\documentclass[12pt]{article}
\usepackage{amsmath}
\usepackage{amssymb}
\usepackage{epsfig}
\usepackage{cite}
\usepackage{color,colordvi}

\begin{document}
\vspace{0.01cm}
\begin{center}
{\Large\bf   On the algebraic meaning of quantum gravity for closed Universes} 

\end{center}

\vspace{0.1cm}

\begin{center}

{\bf Cesar Gomez}$^{a}$\footnote{cesar.gomez@uam.es}

\vspace{.6truecm}

{\em $^a$
Instituto de F\'{\i}sica Te\'orica UAM-CSIC\\
Universidad Aut\'onoma de Madrid,
Cantoblanco, 28049 Madrid, Spain}\\

\end{center}


\begin{abstract}
\noindent  
 
{ Assuming the von Neumann algebra associated with a generic de Sitter observer is properly infinite (type III) we use Connes cocycle to identify the unique ( up to unitary equivalence) background independent dominant weight on an extended algebra formally including an external observer. The background independent algebra defined as the centralizer of this dominant weight is equipped with a semi finite trace. Quantum gravity effects, as well as an algebraic version of the no boundary Hartle Hawking partition function, can be defined by representing, classically different de Sitter closed Universes, as equivalence classes of finite projections in the background independent algebra. The associated generalized entropy is determined by the value of the semi finite trace on each equivalence class of finite projections.

}

\end{abstract}

\thispagestyle{empty}
\clearpage
\tableofcontents
\newpage
\section{Introduction and summary}

The goal of this note is to propose a simple algebraic approach to quantum gravity. We will define, within this algebraic scheme, a quantum gravity Hartle-Hawking type of wave function on an ensemble of different de Sitter closed Universes \footnote{For a recent suggestion on the algebraic version of Hartle-Hawking see \cite{Witten0}.}. We will show that background independence uniquely fixes, up to unitary equivalence, this quantum gravity wave function in terms of a dominant weight $\bar\omega$ on a {\it properly infinite} type $III$ von Neumann algebra associated to the static patch of a generic observer. In this algebraic approach different classical de Sitter Universes are associated to different equivalence classes $[P]_{\alpha}$ of finite projections. The generalized Bekenstein-Gibbons-Hawking \cite{Bek,Bek2,GH} entropy associated to a classical de Sitter Universe $[P]_{\alpha}$ is determined, up to a global rescaling, by the value of $\ln(\bar\omega([P]_{\alpha}))$ for the dominant weight $\bar\omega$ fixed by background independence.  In this introduction we will present, in qualitative terms, the underlying logic flow of the construction as well as some basic definitions.\footnote{The mathematical framework in which we will work will be based on the algebraic approach to de Sitter initiated in \cite{Witten1}. The reader could find most of the mathematical tools used in this note in the reviews \cite{Jones} and  \cite{Wittenrev}. For some recent papers discussing different uses of type $III$ von Neumann algebras in the context of quantum gravity see among others \cite{many1,many2,many3,many4,many5,many6,many6.1,many7,many8,many9,many10,many11,many12,many13}.}

In what follows we will consider a generic observer in de Sitter and we will assume that associated to the causal domain of the observer i.e. the corresponding {\it static patch}, we can define a $C^*$ algebra $A$ of observables. We will think of $A$ as representing all physical properties  the observer can actually measure performing local experiments in her causal domain. We will assume that the observer can use some QFT operator product expansion OPE to identify the multiplication rule defining the algebra $A$.\footnote{For the definition and characterization of observers in the context of QFT in curved space-time see \cite{Witten-o,Witten-2o} and for the role of OPE see \cite{Witten0}. For an early discussion of observers in closed universes and its relation to the notion of reference frames \cite{AS},\cite{RF}, see \cite{Page} and \cite{many6.1}. For the connection of observers and clocks see \cite{Witten0},\cite{many7} and also discussion in \cite{Susskind1}.} 

The concrete relation between the abstract algebra $A$ and the results of the actual observations performed by the observer will be represented by a positive linear form $\phi$ on $A$ to which we will refer as a {\it background}.\footnote{Although these positive linear forms are known as {\it states} we will find more intuitive to denote them as backgrounds. The reasons will become clear along the text.}  

This linear form allows us to define a {\it representation} ( using the GNS construction ) of $A$ as a subalgebra of bounded operators acting on a background dependent Hilbert space $H_{\phi}$. Using these data, namely $A$ and the linear form $\phi$, we can represent the expectation value we observe when measuring a given property $a\in A$ ( in the observer's causal domain ) as $\phi(a)$. For $\phi$ defining a cyclic representation $\phi(a)$ becomes the familiar expectation value $\langle a\Omega,\Omega\rangle$ with the cyclic state $|\Omega\rangle$ playing the formal role of a QFT vacuum state.

In this formal algebraic approach we should keep in mind that elements in $A$ represent those physical properties that can be locally measured in the observer's causal domain and that the values $\phi(a)$ intend to represent the actual expectation values obtained in those local experiments. In the case of de Sitter, where the causal domain is a bounded region of space time, while $A$ represents the algebra of properties that can be measured in the observer's causal domain, the Hilbert space $H_{\phi}$, on which the GNS representation of $A$ is defined, represents quantum states describing, in principle, regions of space time that are not causally connected with the observer's causal domain \footnote{This will depend on the type of the von Neumann algebra of operators obtained in the GNS representation. In this note we will assume that this algebra is type $III$.}. For instance the commutant $A^{'}$ for the representation of $A$ as acting on $H_{\phi}$ defines, for certain linear forms $\phi$,\footnote{In particular those satisfying the trace property.} local observables with support in the complementary static patch.\footnote{These statements will be defined rigorously in the text. In this introduction we simply try to stress that although the algebra $A$ is associated to the local measurements that can be performed by the observer and in that sense only accounts for the information that is  physically accessible to the observer, the quantum states in the Hilbert space where this algebra is acting, once we define a representation using, for instance, the GNS construction for a given background $\phi$, describe regions that are not accessible to the observer.}
Thus we should a priori expect that $\phi(a)$ will account for any sort of Unruh \cite{Unruh} like thermality, reflecting the ignorance of the observer about what takes place beyond her causal domain.

Before going into a more detailed discussion on how to define the basic data $A$ and $\phi$ we need to understand how we can use {\it actual observations described by elements in $A$ to identify a particular quantum state in $H_{\phi}$}. In quantum mechanics it is standard to identify a quantum vector state, let us say $|\psi\rangle$, in $H_{\phi}$ with a {\it minimal projection} $P_{\psi}$. Physically this projection can represent an observable if we can design, in the causal domain, a Yes/No {\it local experiment} that can be represented by the minimal projection i.e. if $P_{\psi}$ is in the algebra $A$. Thus, the first crucial question is: Do we have in $A$ minimal projections? In other words: Can we design local experiments in the causal domain defined by the static patch, that select a quantum state?

An important peculiarity of local QFT when we work with a bounded region of space-time, as it is the static patch, is that the associated algebra $A$ describing the accessible local information \footnote{Defined as acting on the Hilbert space $H_{\phi}$.}, does not have any non zero {\it finite projection} \footnote{Given two projections $P$ and $Q$ in $B(H)$ for $H$ a Hilbert space, we say that $P$ is finite if $P\leq Q$ (meaning $PH \subseteq QH$) and $P\sim Q$ ( for $\sim$ meaning equivalent projections) implies $P=Q$. Two different projections $P$ and $Q$ are equivalent if there exist a partial isometry $u$ such that $P=u^*u$ and $Q=uu^*$.}. An algebra with this property is known as {\it properly infinite} \footnote{For an intuitive physics review see \cite{Ingvason} and references therein.}.

This will be our basic assumption, namely that for any background $\phi$ the algebra $A$ acting on the Hilbert space $H_{\phi}$ associated with the static patch is {\it properly infinite} or type $III$.\footnote{Note we are talking about the von Neumann algebra acting on $H_{\phi}$.} This is an assumption that derives from locality of the QFT and on the existence of a horizon that makes the causal domain of a generic observer a bounded region of space-time. The non existence of finite projections in $A$ is an enormously strong condition that, as said, means that the observer cannot design any local experiment
able to select a finite dimensional subspace of the Hilbert space of states. This fact also explains the importance of traces. Intuitively you can define a finite projection as one with finite standard trace \footnote{This is just an intuitive definition. As we will see we can define traces with finite values that cannot be identified with the standard trace.}, so since in $A$ we don't have any finite projection we can also conclude that we don't have any well defined finite trace. 

Once we have assumed that for the case of de Sitter and for any observer, the algebra 
$A$, as acting on $H_{\phi}$ for any {\it background} $\phi$, is properly infinite \footnote{From now on whenever we refer to the algebra $A$ we will assume is properly infinite.},  we need to figure out, in order to couple to gravity, how to implement general covariance. We will define {\it general covariance} as {\it background independence} i.e. as independence on the particular background $\phi$ we use to represent the algebra as an algebra of operators acting on a Hilbert space.

In order to identify the dependence on the background $\phi$ we need to {\it compare} the representations of $A$ defined by different backgrounds. This can be done using Connes cocycle \cite{Connes1}. Given two backgrounds for the algebra $A$, Connes cocycle defines a one parameter map $t\rightarrow u_t$ with $u_t$ unitary elements of $A$, relating the associated representations of $A$. Two backgrounds are physically equivalent if they are related by a unitary operator $U$ in $A$. Thus, in order to define background independence, we need to extend the algebra $A$ into some bigger algebra, let us say $A_{ext}$, in such a way that the cocycle {\it map} $t\rightarrow u_t$ associated to two arbitrary backgrounds on $A$ is represented by a unitary operator $U$ in $A_{ext}$. In other words we need to map the cocycle $t\rightarrow u_t$ with $u_t$ a one parameter family of unitary elements of $A$ into a unitary operator $U$ in the extended algebra. Once we have identified the extended algebra and the way to define $U$ we can achieve background independence \footnote{In the sense of unitary equivalence of any two backgrounds on the original algebra $A$.} in the extended algebra. The underlying logic of this construction is very familiar in our physics approach to gauge invariance. In that case we add some degrees of freedom and we define gauge invariant quantities in the extended system. The analog in our case will be to define for each background $\phi$ on $A$ a "gauge invariant" background $\omega(\phi)$ on the extended algebra $A_{ext}$ such that $\omega(\phi)$ for different $\phi$'s are {\it unitarily equivalent}.

Interestingly enough the solution to this {\it background independence} problem has a {\it unique} solution when $A$ is properly infinite. The solution \cite{Connes1},\cite{Connes2} is that $A_{ext}=A\otimes F_{\infty}$ for $F_{\infty}$ the algebra of bounded operators on $L^2(\mathbb{R})$ and the unique "gauge invariant background" $\omega(\phi)$ is the {\it dominant weight} $\bar\omega=\phi\otimes \omega$ with $\omega$ the weight on $F_{\infty}$ defined by $Tr(x\rho)$ with $\rho$ the generator of translations in $L^2(\mathbb{R})$ and $x$ any element in $F_{\infty}$. Although at first sight this answer can look too formal and abstract it already identifies the key ingredients needed to achieve background independence. Namely the added "degrees of freedom" are the ones we represent by the algebra $F_{\infty}$ that is simply the full algebra of bounded operators acting on a quantum system of one degree of freedom with Hilbert space $L^2(\mathbb{R})$. This extra system is what we can identify with the origin of {\it external observer}. Secondly it implies that we need to work with weights since the trace in $F_{\infty}$ used to define $\omega$ can take values in the full interval $[0,\infty]$ \footnote{We will introduce the definitions of weights in the next section. At the level of this qualitative introduction we only want to stress that the background $\bar\omega$ achieving background independence is a linear map that can take values in $[0,\infty]$.}.

Now, as done in the case of gauge theories, we need to identify, in the extended algebra $A\otimes F_{\infty}$, the {\it background independent subalgebra}. This subalgebra should be uniquely determined by the background independent weight $\bar\omega$ and can be identified as the centralizer ${\cal{C}}_{\bar\omega}$ that is acting on the extended Hilbert space ${\cal{H}}=:L^2(\mathbb{R},H_{\phi})$ for $\phi$ any background on $A$. The logic for this choice is that due to the KMS property for $\bar\omega$ the centralizer admits a weight semi-finite trace, defined by $\bar\omega$ taking values in $[0,\infty]$. 

The algebra ${\cal{C}}_{\bar\omega}$ defines the "gauge invariant" (background independent ) algebra of observables and implements the needed "gauge dressing". This background independent algebra is mathematically interesting for two reasons whose meaning  we will briefly discuss in the text. One reason is that the algebra has a non trivial center, the other reason is the existence of an automorphism of the algebra that implements global scale transformations $\bar\omega \rightarrow \lambda\bar\omega$ for $\lambda >0$.

The final question we need to answer in order to get some glimpses of quantum gravity is how classically different de Sitter space-times, for instance with different values of the cosmological constant, are embedded in this extended algebra. The existence of a trace on the gauge invariant ( background independent) algebra ${\cal{C}}_{\bar\omega}$ \footnote{See \cite{Connes2} and \cite{Takesaki}}, allows us to find background independent projections $P$ in ${\cal{C}}_{\bar\omega}$ with finite trace and to identify background independent subalgebras as $P{\cal{C}}_{\bar\omega}P$.  We will identify the different de Sitter space-times with equivalence classes $[P]_{\alpha}$ of finite projections\footnote{Recall that two different projections $P_1$ and $P_2$ are equivalent if there exist a partial isometry $u$ such that $P_1=u^*u$ and $P_2=uu^*$.}. The interest of the meaning of this equivalence relation is that equivalent projections have the same value of the trace. Thus the trace associated to $\bar\omega$ defines a map $[P]\rightarrow \bar\omega([P])$ from the set of equivalence classes, that we are identifying as representing different de Sitter space time Universes, into the interval $[0,\infty]$. We will identify the logarithm of this trace as representing the generalized Gibbons Hawking entropy $S_{\alpha}$ of the de Sitter Universe associated to the equivalence class $[P]_{\alpha}$ as \footnote{The relation of this definition of entropy with a von Neumann type of entropy for some {\it affiliated density matrix} will be discussed in section 7.}
\begin{equation}\label{one}
S_{\alpha}\sim \ln(\bar\omega([P]_{\alpha})
\end{equation}

Note that each de Sitter Universe is associated with the background independent subalgebra $P{\cal{C}}_{\bar\omega}P$ for $P$ any finite projection in the corresponding equivalence class.
These subalgebras are the type $II_1$ algebras identified in \cite{Witten1}.

An important comment to be further discussed in section 7 is to alert the reader on the meaning of the finiteness of the projections, used to define the representation of classically de Sitter space times, in the background independent algebra ${\cal{C}}_{\bar\omega}$. The microscopic meaning of the finite value of $\bar\omega(P)$ entering into (\ref{one}) has been the subject of many discussions in the past. In \cite{Banks4} this finiteness was interpreted as indicating a finite dimensional Hilbert space. The algebraic approach developed in this note sheds some light on this issue. In particular the finiteness of the projection and consequently of the generalized entropy defined in (\ref{one}) by no means implies the finite dimension of the Hilbert space $P{\cal{H}}$. As we will see in section 7 the type $II_1$ factor $P{\cal{C}}_{\bar\omega}P$ for $P$ a {\it finite} projection contains subalgebras isomorphic to $M_n(\mathbb{C})$ for any natural number $n$. In other words contains subalgebras representing any arbitrary number of q-bits.

Finally we can compare the different de Sitter space times associated to different equivalence classes of projections. With each equivalence class of projections $[P]$ we will associate a maximal entropy vector state $\Psi_{[P]}$ \footnote{More specifically a square summable sequence of vector states $\Psi_{[P]}^i$ in $l^2({\mathbb{N}},{cal{H}})$.} the Hilbert space ${\cal{H}}$ representing ${\cal{C}}_{\bar\omega}$  and we will define quantum amplitudes between different dS Universes as $\langle\Psi_{[P]},\Psi_{[P']}\rangle$ with the scalar product the one defining the Hilbert space representation of ${\cal{C}}_{\bar\omega}$.

 The existence of a one parameter automorphism $\theta_s$ of ${\cal{C}}_{\bar\omega}$ scaling the weight $\bar\omega$ induces a "scale transformation" on the space of equivalence classes of finite projections, let us say $[P]\rightarrow [P]_s$. This is, in our representation, a transformation ( flow) on the space of classically different de Sitter space-times. We can heuristically think of this scale transformation as a sort of {\it de Sitter renormalization group flow}. By analogy with the definition of the {\it no boundary} Hartle Hawking euclidean wave function \cite{HH} we will define a formal Hartle-Hawking weight as {\it the scale invariant weight} \footnote{Where $\cdot$ represents any element in ${\cal{C}}_{\bar\omega}$ and $[P]_s(\cdot)$ represents $P_s\cdot P_s$ i.e. an element in $P_s{\cal{C}}_{\bar\omega}P_s$.}.  
\begin{equation}\label{two}
\bar\omega_{HH}(\cdot)=\int_s\bar\omega([P]_s(\cdot))ds
\end{equation}
where the integral over the de Sitter flow parameter $s$ is on the interval $\lambda=e^{-s}>0$ and where (\ref{two}) is defined for an arbitrary "reference" finite projection $P$. The careful study of this integration goes beyond the limited scope of this note.
The {\it no boundary quantum gravity} HH partition function $Z$ will be defined as $Z=\bar\omega_{HH}(I)$ for $I$ the identity of the background independent algebra ${\cal{C}}_{\bar\omega}$.

Summarizing the logic flow goes as follows:

\begin{itemize}
\item LQFT and bounded causal domain of the observer $\Rightarrow$ properly infinite von Neumann algebra $A$
\item Background independence $\Rightarrow$ the existence of a unique dominant weight $\bar\omega$ on $A\otimes F_{\infty}$
\item KMS $\Rightarrow$ the centralizer ${\cal{C}}_{\bar\omega}$ as the background independent algebra and the existence, in this algebra, of finite projections.
\item Equivalence classes $[P]$ of finite projections in ${\cal{C}}_{\bar\omega}$ define different physical de Sitter space-times, associated to the subalgebras $P{\cal{C}}_{\bar\omega}P$
\item The formal {\it no boundary} Hartle Hawking (HH) weight is defined by integrating on the action of the scale transformation automorphism of ${\cal{C}}_{\bar\omega}$ on the space of equivalent finite projections

\end{itemize}

The paper is organized as follows. In section 2 we review some basics on the GNS construction, the definition of weights and the KMS property. In section 3 we basically use Lemma 1.2.5 in \cite{Connes1} to define background independence. We define the extended algebra, that can be interpreted as including the external observer, and we identify the background independent dominant weight. In section 4 we work out the background independent algebra ${\cal{C}}_{\bar\omega}$ and we define the trace and the dressing. We briefly discuss the connection of dressing and conditional expectations as well as the affiliated density matrices. In section 5 we briefly discuss\footnote{Following \cite{Witten0}.} the role of OPE and the smearing of short distances divergences in the definition of the background independent algebra. Section 6, that can be avoided in a first reading addresses the relation of the added external observer to physical clocks and finite projections. In section 7 we define the quantum gravity superpositions for an ensemble of different de Sitter space times in terms of the background independent dominant weight \footnote{More precisely in terms of the trace defined by the weight dual to the dominant weight \cite{Takesaki}.} and we discuss the connection to the no boundary HH partition function.

\section{Observables and backgrounds: some preliminary comments}
In Quantum Mechanics as well as in QFT  we can define the abstract notion of a $C^{*}$ algebra of observables. Recall that a $C^{*}$ algebra $A$ is an algebra with an involution $(*)$ satisfying the properties of the {\it adjoint} and a norm $||a||$ for $a\in A$ that defines on $A$ a notion of distance. We can always assume that $A$ contains the identity $I\in A$. In this algebra physical observables are defined as the self adjoint elements i.e. $a^*=a$. The crucial and less obvious ingredients in this definition are the norm that makes the algebra $A$ a Banach space as well as the physics underlying the definition of the {\it multiplication rule}. 

In doing physics we are interested in defining positive complex valued linear forms $\phi:A\rightarrow {\mathbb{C}}$ with $\phi(a^*a)\geq 0$ and such that $\phi$ is continuous with respect to the norm of $A$. Since $\phi(a^*) =\overline{\phi(a)}$ this linear form assigns to any self adjoint observable, a real number that we can physically identify with the {\it expectation value} we get experimentally by performing local measurements. In that sense $\phi$ encodes information both on the actual observer that is measuring as well as on the particular state of the observed system. The continuity of $\phi$ reflects the basic intuition that if we consider a sequence of elements $a_n$ of $A$ with a given limit $a$ the results of the actual measuremets $\phi(a_n)$ should have as limit $\phi(a)$.

Until this point we have not assumed, in defining $A$, that this $C^{*}$ algebra is represented as an algebra of bounded operators acting on a Hilbert space of quantum states. A basic result of the theory of $C^{*}$ algebras, known as the GNS construction, associates with a given $C^{*}$ algebra $A$ and a positive linear form $\phi$, such that $\phi(I)=1$, a representation $\pi_{\phi}$ of $A$ into the algebra of bounded operators acting on a Hilbert space $H_{\phi}$. Hence, it defines a {\it representation} of a  $C^{*}$ algebra $A$ as an algebra of bounded operators acting on a Hilbert space. The way to do it is to associate with any element $a$ in $A$ a formal vector state $|a\rangle$ and to define the scalar product as $\langle a,b\rangle = \phi(a^{*}b)$ on the quotient space of equivalence classes $A/N$ with $N$ those elements $x\in A$ such that $|x\rangle$ is a null vector i.e. those such that $\phi(x^{*}x)=0$. After completion this defines the GNS Hilbert space $H_{\phi}$ and the representation $\pi_{\phi}$ of the $C^{*}$ algebra $A$ into $B(H_{\phi})$ by left multiplication $\pi_{\phi}(a) |b\rangle = |ab\rangle$. In these conditions we get $\phi(a)=\langle a \Omega_{\phi},\Omega_{\phi}\rangle$ for $|\Omega_{\phi}\rangle$ the state corresponding, in the GNS construction defined by $\phi$, to the identity i.e. $|\Omega_{\phi}\rangle=|I\rangle$. Note that the state $\Omega_{\phi}$ is cyclic in the sense that $(A/N)|\Omega_{\phi}\rangle$ is dense in $H_{\phi}$. The linear form $\phi$ is called a {\it state} and the corresponding GNS representation is a cyclic representation. On $A/N$ the state $\phi$ is {\it faithful} i.e.  $\phi(x^*x)=0$ implies that $x=0$. 

At this point it is important to stress several properties of the GNS construction. First of all, once we have defined the representation of the $C^{*}$ algebra $A$ as an algebra acting on $H_{\phi}$, for a given ${\phi}$, we can define its commutant $(\pi_{\phi}(A))^{'}$ i.e. the set of bounded operators in $H_{\phi}$ commuting with the algebra $\pi_{\phi}(A(D))$. It is easy to see that the representation $\pi_{\phi}$ is {\it irreducible} if the corresponding commutant is ${\mathbb{C}}I$. States $\phi$ defining irreducible representations will be called {\it pure states}. Once we define the representation of $A$ on $H_{\phi}$ the so defined algebra will be a von Neumann algebra if it satisfies the bicommutant relation 
$(\pi_{\phi}(A))^{''}= (\pi_{\phi}(A))$. 

Many times in physics we are interested in faithful states that satisfy the trace property. In order to start appreciating the importance of this property notice that if the linear form $\phi$ satisfies the trace property, namely
\begin{equation}\label{tracedef}
\phi(ab)=\phi(ba)
\end{equation} 
then the commutant $(\pi_{\phi}(A))^{'}$ is isomorphic to $A$ and acting on $H_{\phi}$ by {\it right multiplication}. In particular in this {\it trace case} we can define an antilinear {\it isometry} $J_{\phi}$ such that 
\begin{equation}\label{anti}
(\pi_{\phi}(A))^{'}=J_{\phi}\pi_{\phi}(A)J_{\phi}
\end{equation}
with $J_{\phi}^2=1$.
This situation is physically specially interesting. Indeed if we characterize the {\it observer}
in terms of the linear form $\phi$ we note that when this linear form is a {\it trace} in the sense of (\ref{tracedef}) the representation of the $C^{*}$ algebra $A$ on the GNS Hilbert space $H_{\phi}$ has a {\it symmetric} commutant isomorphic to $A$ as well as a map $J_{\phi}$ satisfying (\ref{anti}). Note that a state $\phi$ satisfying the trace property is {\it not a pure state} in the sense defined above i.e. the representation $\pi_{\phi}$ is not irreducible.\footnote{The representation $\pi_{\phi}$ in $H_{\phi}$ for $\phi$ a trace state defines the standard form of a type $II_1$ von Neumann algebra. In this case $A$ and the commutant are perfectly symmetric. See \cite{Jones} chapter 6.}

Something that can sounds at first sight confusing is that the state defined by the linear form $\phi$ can be reducible i.e. not pure, while the corresponding cyclic vector state $|\Omega_{\phi}\rangle$ in $H_{\phi}$ is a {\it pure vector state} in $H_{\phi}$. This phenomena is very familiar in physics and it is known as {\it purification}. In words the cyclic state $|\Omega_{\phi}\rangle$ defines a {\it purification} of the linear form $\phi$ i.e. $\phi(a)=\langle a\Omega_{\phi},\Omega_{\phi}\rangle$. A simple example is the TFD pure state as a purification of a linear form $\phi$ defining thermal correlators. 

For a given $\phi$ we can interpret the cyclic state $|\Omega_{\phi}\rangle$ as representing a {\it background configuration} and the GNS Hilbert space $H_{\phi}$ as describing, in a sort of Fock representation, {\it perturbative quantum fluctuations} around  this background. In this sense we can say that two different states $\phi_1$ and $\phi_2$ represent the same background if they are {\it unitarily equivalent} i.e. if there exist a unitary transformation $U$ in $A$ between $H_{\phi_1}$ and $H_{\phi_2}$.  We can also define a {\it universal} representation where we consider the direct sum of all not equivalent cyclic representations. 

In order to make the former "background" picture a bit more intuitive let us briefly recall a familiar example in the context of holography \footnote{This example is only used to provide some intuition based on a familiar case. In this note we will not consider the case of eternal AdS black holes that is the physics underlying the following example.}.

\subsection{Semi-classical states}
In the holographic frame based on the AdS/CFT correspondence  the boundary CFT can be characterized by the corresponding OPE. This OPE can be in principle used to equip the full set of local boundary operators with the structure of a $C^{*}$ algebra. As discussed in \cite{many2} if we consider, for instance in $N=4$ super Yang Mills, the set of single trace operators and we appropriately truncate the OPE by taking the large $N$ limit of the OPE coefficients, we can define, at large $N$, a $C^*$ algebra $A_{\infty}$ ( see \cite{many2} and references therein for a more detailed description). Correlators among these operators with a well defined large $N$ limit define faithful {\it states} $\phi$ on $A_{\infty}$ i.e. continuous linear forms on $A_{\infty}$ such that $\phi(x^*x)=0$ implies $x=0$. As discussed, these states define a GNS {\it representation} $\pi_{\phi}$ of the large $N$ $C^*$ algebra $A_{\infty}$ in a {\it state dependent} Hilbert space $H_{\phi}$ with a cyclic state $\Omega_{\phi}$. In the large $N$ expansion this Hilbert space describes $O(1)$ small fluctuations around a semiclassical background defined by the semiclassical state $\Omega_{\phi}$. Two different semiclassical states $\phi_1$ and $\phi_2$ define different Hilbert spaces $H_{\phi_1}$ and $H_{\phi_2}$ as well as different cyclic states $\Omega_{\phi_1}$ and $\Omega_{\phi_2}$. In order to relate the perturbative large $N$ dynamics on two different semiclassical states we will need to define some operator $U_{1,2}:H_{\phi_1}\rightarrow H_{\phi_2}$ such that $U_{1,2}|\Omega_{\phi_1}\rangle=|\Omega_{\phi_2}\rangle$. Obviously we observe that if $U_{1,2}$ is {\it unitary} both semiclassical states are physically identical i.e. they define the same correlators \footnote{This means that $\phi_1(O)=\phi_2(O)$ for any $O\in {\cal{A}}$.}. In other words, two {\it physically different} semiclassical states lead to two representations of the OPE $C^*$ algebra that are {\it not} unitarily equivalent. Intuitively we could imagine that once we embed the large $N$ algebra into the full algebra, defined by the finite $N$ OPE, the large $N$ Hilbert spaces $H_{\phi}$ associated to semiclassical states $\phi$ will define different disconnected Hilbert subspaces and that the operator $U_{1,2}$  between different subspaces will necessarily involve $\frac{1}{N}$ effects. Note that these effects will correspond to $\frac{1}{G}$ effects in the holographic dual. In this note we will try to define a concrete mathematical framework where some of these intuitions could be made manifest for the case of de Sitter closed universes. The extension to the large $N$ holographic cases will not be discussed in this note.

\subsection{Weights}
As we will see the notion of weight ( se \cite{Takesakibook}) is very useful to address the problem of background independence. Before going into details let us motivate the definition of weights.
 
Starting with a state $\phi$ we have defined, using the GNS construction, the cyclic representation $\pi_{\phi}$ of $A$ as a von Neumann algebra acting on the Hilbert space $H_{\phi}$. Let us denote $A^{+}$ the set of elements in $A$ that are of the form $a^*a$. The state $\phi$ acting on $A^{+}$ associates to any element $a^*a$ the {\it norm} of the vector in $H_{\phi}$ associated to $a$ i.e. $\phi(a^*a)= \langle a,a\rangle$. Thus if the states in $H_{\phi}$ have finite norm the state $\phi$ on $A^{+}$ defines a positive and finite linear form. A weight $\omega$ is defined as a linear map on $A^{+}$ satisfying $\omega(x+y)=\omega(x)+\omega(y)$ and $\omega(\lambda x)=\lambda\omega(x)$ taking values in $[0,\infty]$. Note that since now the value of the weight $\omega(a^*a)$ can be $\infty$ we cannot represent this value as the norm of the vector in $H_{\phi}$ associated to $a$.

In what follows we will normally work with {\it semi-finite} weights. For a weight $\omega$ we can define $A_{\omega}=\{ x\in A^+; \omega(x)<\infty\}$. We will say that the weight is semi-finite if $A_{\omega}$ generates the full algebra $A$. We will also use faithful weights i.e. those such that $\omega(x)=0$ implies $x=0$ and normal where by that we mean that they satisfy $\omega(\sup x_i)=\sup(\omega(x_i))$ for every increasing and bounded set of $x_i$ in $A^{+}$. 

In order to appreciate the physics meaning of weights imagine 
a weight  $\omega$ that satisfies the trace property, namely $\omega(a^*a)=\omega(aa^*)$. In these conditions you can use this weight to define a semi-finite trace $Tr$ on $A^+$ i.e. a trace that can take values in $[0,\infty]$ and where the trace of the identity $Tr(I)$ is infinity. Thus, in this case, although we can find a weight to define $Tr(a)$ as $\omega(a)$ we cannot {\it purify} this trace as $\langle a \Psi,\Psi\rangle$ for some normalizable state $|\Psi\rangle$ in the Hilbert space $H_{\phi}$. In other words weights cannot be purified. \footnote{As a simple example in quantum mechanics, that we will extensively use along the note, consider the type $I_{\infty}$ algebra $F_{\infty}$ defined by the bounded operators acting on the Hilbert space $L^2(\mathbb{R})$. In this case you cannot define a finite trace on $F_{\infty}$ but you can define weights leading to semi-finite traces.}

\subsection{The surprising appearance of thermodynamics: the KMS condition}
Recall that we started with the $C^{*}$ algebra $A$ and we used a positive linear form $\phi$, that we will assume faithful and normal, to define, using the GNS construction, a representation $\pi_{\phi}$ of $A$ on the algebra of bounded operators acting on the Hilbert space $H_{\phi}$ with cyclic and separating state $|\Omega_{\phi}\rangle$. From now on we will denote also $A$ the algebra $\pi_{\phi}(A)$ and we will assume that the so defined algebra is a von Neumann algebra. Given these data , namely $H_{\phi}$ and $|\Omega_{\phi}\rangle$, we can define the one parameter {\it modular automorphism} $\sigma_{\phi}^t$. This automorphism is defined by first lifting the action of the adjoint $*$ operation to $H_{\phi}$. This defines the {\it unbounded} Tomita-Takesaki operators $S_{\phi}$. Secondly we use {\it the polar decomposition} to represent $S_{\phi}= J_{\phi}\Delta_{\phi}^{1/2}$, and finally we define the modular automorphism $\sigma_{\phi}^t(a)=\Delta_{\phi}^{it}a\Delta_{\phi}^{-it}$ for $a\in A$. 

For $A$ a von Neumann algebra the {\it type} of this algebra is partially determined by the properties of the modular automorphism. In particular if $A$ is a factor i.e. a von Neumann algebra with trivial center, and the modular automorphism $\sigma_{\phi}^t$ is {\it outer} i.e. not part of the algebra $A$, for any $\phi$ and any $t$ then the algebra $A$ is a type $III$ factor. It is easy to check that for the particular case where the linear form $\phi$ satisfies the trace property we get $S_{\phi}=J_{\phi}$ with $J_{\phi}$, the antilinear map used in (\ref{anti}). Thus in this case the modular automorphism is trivially inner. This already implies that for type $III$ factors {\it we cannot find a state $\phi$ satisfying the trace property}. Another way to characterize type $III$ algebras is in terms of the properties of the projections. Type $III$ algebras are {\it properly infinite} in the sense that there does not exist any non zero finite projection $P$ \footnote{For instance in the case of the type $I_{\infty}$ algebra defined by the algebra of bounded operators on a separable Hilbert space $H$ we have obviously finite projections as well as minimal projections associated to any one dimensional subspace of $H$. For a type $II_1$ algebra $A$ acting on a Hilbert space $H$ we have a well defined finite trace $tr$ and finite projections $P$ in $A$ with $tr(P)$ finite, but not minimal projections. In the semi-finite case of $A$ being a type $II_{\infty}$ algebra we have a semi finite trace $Tr$ and  some finite projections $P$ such that $Tr(P)$ is finite. In the properly infinite case there is not any trace linear form such that $Tr(P)$ is finite for $P$ any projection.}. In the physics applications we will discuss in this note we will assume the von Neumann algebra $A$ is {properly infinite} i.e of type $III$ \footnote{A different way to characterize a type $III$ factor $A$ is considering the {\it spectrum} of the modular Hamiltonians $h_{\phi}$ i.e. the generator of $\Delta_{\phi}$. Denoting $S(A)$ the intersection of the spectrum of $h_{\phi}$ for all faithful and normal states $\phi$ the factor $A$ will be type $III$ iff $0\in S(A)$. The case of $III_1$ factors that we will mostly consider in this note are characterized by $S(A)={\mathbb{R}}^{+}$ i.e. the modular Hamiltonian has continuous spectrum}. 

The key property of the modular automorphism is that it is the {\it unique} automorphism relative to which the state $\phi$ satisfies the KMS condition \footnote{The interested reader can find the proof of this basic theorem in the Appendix B of \cite{Jones}.}. This condition implies that
\begin{equation}\label{inv}
\phi (\sigma_{\phi}^t(a))=\phi(a)
\end{equation}
for $a\in A$ and the existence, for each couple of elements $x$ and $y$ in $A$, of a function $F(t)$, analytic on the strip $\{t; 0\leq Im t \leq 1\}$, and such that $F(t)=\phi(\sigma_{\phi}^t x y)$ and $F(t+i)=\phi(y\sigma_{\phi}^t x)$. More explicitly we can write the KMS condition as
\begin{equation}\label{kmsc}
\phi(y \sigma_{\phi}^t (x))=\phi(x \sigma_{\phi}^{t+i} (y))
\end{equation}
where recall $\phi (xy)=\langle x y \Omega_{\phi},\Omega_{\phi}\rangle$. 

The KMS condition (\ref{kmsc}) is very familiar in statistical mechanics ( see \cite{haag} and discussion in \cite{roveli}). Indeed, for thermal correlators defined as $\langle AB\rangle_{\beta}= Tr(\rho_{\beta} AB)$ with $\rho_{\beta}$ the Gibbs canonical distribution $\rho=e^{-\beta \hat H}$ for a positive Hamiltonian $\hat H$, we can prove the KMS relation
\begin{equation}\label{kmsc2}
\langle A(0)B(t)\rangle_{\beta} = \langle B(0) A(t+i\beta)\rangle_{\beta}
\end{equation}
for the standard time automorphism $A(t)=e^{i\hat Ht} A(0)e^{-i\hat H t}$ defined by the Hamiltonian $\hat H$. Comparing (\ref{kmsc}) and (\ref{kmsc2}) we observe that the linear form $\phi$ defines "correlators" satisfying the KMS property at a formal temperature $\beta$ if we use the generator of the modular automorphisms $\Delta_{\phi}$ i.e. the modular Hamiltonian $h_{\phi}$, as $-\beta\hat H$ \footnote{Note already that if $\phi$ satisfies the trace property the KMS condition (\ref{kmsc}) becomes
\begin{equation}\label{kmsm}
\phi(y \sigma_{\phi}^t (x))=\phi(x \sigma_{\phi}^{t} (y))
\end{equation}
that naively could be interpreted as describing thermal correlators at infinite temperature $\beta=0$. See for instance \cite{Banks9} and discussion in \cite{Susskind2}.}. 

It is important to stress that although the KMS relation for the state $\phi$ relative to the modular automorphism $\sigma_{\phi}^t$ looks formally as the standard KMS condition for thermal correlators this does not implies the existence of a representation of $\phi(xy)$ as $tr(\rho_{\beta} x y)$ for some well defined and self adjoint density matrix $\rho_{\beta}$. Indeed for a type $III$ algebra such a representation does not exists although $\phi(xy)$ satisfies the KMS condition (\ref{kmsc}) for $\sigma_{\phi}^t$.

In this note we will be interested in identifying the von Neumann algebra $A$ acting on $H_{\phi}$ with the type $III_1$ factor we associate to the static patch of a generic observer in de Sitter. In this case the KMS condition for $\phi$ is telling us that the expectation values defined by the state $\phi$ i.e. $\phi(a)$ for any $a\in A$ satisfies the standard KMS property but, neither the existence of a canonical Gibbs density matrix $\rho_{\beta}$ nor the existence of a trace state \footnote{The formal KMS thermality of the correlators defined by any faithful and normal state $\phi$ on the type $III_1$ factor associated with the static patch could be naively interpreted as a proof of the familiar de Sitter {\it Gibbons-Hawking} thermality.
However this formal KMS thermality associated with the modular automorphism is not the familiar thermality associated to the Euclidean version of de Sitter that you will try to define in terms of a positive physical Hamiltonian.}.
 
{\bf Remark} When doing standard Statistical Mechanics we consider the algebra $A$ of observables as acting on a Hilbert space of states and we define the thermal correlators as $Tr(a\rho)$ for $a$ any element in $A$ and for $\rho$ the density matrix operator also in $A$ defining the corresponding canonical ensemble. The trace used in this definition is defined on the whole algebra $A$ and it is assumed to be finite. The so called TFD purification of the density matrix $\rho$ is defined by a pure state $|TFD_{\rho}\rangle$ in a formally doubled Hilbert space satisfying $\langle TFD_{\rho}, a TFD_{\rho}\rangle = Tr(a\rho)$. In the abstract algebraic setup described above we get the standard Statistical Mechanics picture whenever we work with an algebra $A$ that admits a finite trace i.e. whenever we work with a {\it finite} von Neumann algebra of observables. We will come back to this discussion in section 4 where we will briefly mention the notion of affiliations.

\section{Background independence}
In the previous section we have formally defined, for a given $C^*$ algebra of observables $A$, a {\it background} as any faithful and normal state $\phi$ on $A$. In order to {\it compare} different backgrounds we need to identify a precise mathematical framework where we can put together the different representations of $A$ defined by different states $\phi$. This mathematical frame is defined by Connes cocycle \cite{Connes1}. 

Intuitively, as we will see in a moment, the simplest way to compare two states $\phi_1$ and $\phi_2$ i.e. two backgrounds, will be to extend the algebra $A$ to $A\otimes M_2(\mathbb{C})$ with $M_2(\mathbb{C})$ the algebra of two by two complex matrices \footnote{The reader can find suggestive to interpret the extension to $A\otimes M_2(\mathbb{C})$ as reflecting the inclusion of a couple of two external q-bits.}. In this extended algebra we can identify two copies of $A$ and to act with the different states on each copy. In order to define {\it background independence} we will like first to compare {\it all possible states} ( backgrounds). This as we will discuss, requires to replace $M_2(\mathbb{C})$ ( used to compare two states) by the algebra of "infinite" matrices acting on a separable Hilbert space \footnote{Intuitively we can say that background independence i.e general covariance, requires to "add" an infinite number of q-bits.}. Once we identify this algebra with the $I_{\infty}$ factor $F_{\infty}$ of bounded operators acting on a Hilbert space $L^2(\mathbb{R})$ the question of background independence reduces to identify weights on $A\otimes F_{\infty}$ which are {\it independent} of the concrete background $\phi$ on $A$. In this section we will identify the background independent state on $A\otimes F_{\infty}$ with the unique, up to unitary equivalence, {\it dominant weight} \footnote{This uniqueness refers to the equivalence relation between weights defined in \cite{Connes2}.}. In order to make the presentation self contained we will go through the basic technical details needed to define this background independent {\it weight}.

\subsection{Connes cocycle}
Once we have defined the modular automorphism and the KMS property we can start comparing different faithful and normal states $\phi$.

The first thing we can do to relate two states, let us say $\phi_1$ and $\phi_2$, is to identify the corresponding Connes cocycle $u_{1,2}(t)$.\footnote{See \cite{Connes1}, chapter 14 in \cite{Jones} and \cite{many3}.} Recall that for a given von Neumann algebra $A$ and for two faithful and normal states $\phi_1$,$\phi_2$ Connes cocycle is defined by the continuous map $t\rightarrow u_{1,2}(t)$ from $\mathbb{R}$
into the {\it unitary} group of $A$ by the relation
\begin{equation}
\sigma_1^t=u_{1,2}(t)\sigma_2^t u^{*}_{1,2}(t)
\end{equation}
for ${\sigma_i}^t$ with $i=1,2$ the modular automorphisms associated to the two different states. The  map $t\rightarrow u_{1,2}^t$ is {\it not} satisfying the group composition rule $u_{1,2}^t u_{1,2}^{t'}=u_{1,2}^{t+t'}$ but instead the cocycle condition\footnote{This cocycle makes the "symmetry" transformation relating different backgrounds very close to the recent notion of generalized symmetries.}
\begin{equation}\label{cocycle1}
u_{1,2}^{t+t'}= u_{1,2}^t \sigma_2^t(u_{1,2}^{t'})
\end{equation}
For future use it will be convenient to sketch briefly the derivation of Connes cocycle.

 In order to compare the two states $\phi_1$ and $\phi_2$ on $A$ let us define the extended algebra $A\otimes M_2(\mathbb{C})$ with $M_2(\mathbb{C})$ the algebra of two by two complex matrices. We can visualize $A\otimes M_2(\mathbb{C})$ as two by two matrices valued in $A$. Let us now define a state $\Phi$ on $A\otimes M_2(\mathbb{C})$ as 
 \begin{equation}\label{state}
\Phi((x)_{i,j})= \frac{1}{2}(\phi_1(x_{1,1}) +\phi_2(x_{2,2}))
\end{equation}
 with $(x_{i,j})\in A\otimes M_2(\mathbb{C})$ the matrix valued in $A$. Now recalling our discussion on KMS we observe that the state $\phi_1$ satisfies the KMS property with respect to the modular automorphism $\sigma_{\Phi}^t$ defined by $\Phi$ on $p(A\otimes M_2(\mathbb{C}))p$ for $p$ the projection 
$$
\begin{bmatrix} 
	1 & 0  \\
	0 & 0 \\
\end{bmatrix}
$$
and the same for $\phi_2$ for the projection $q$ defined by
$$
\begin{bmatrix} 
	0 & 0  \\
	0 & 1 \\
\end{bmatrix}
$$
In order to relate both states we can define $V_t=\sigma_{\Phi}^t(1\otimes e_{2,1})$  that satisfies $V_tV^*_t=p$ and $V^*_tV_t=q$.\footnote{The matrices $e_{i,j}$ is a system of unit matrices in $M_2(\mathbb{C})$ i.e. $e_{i,j}$ are two by two matrices with $(i,j)$ the only non vanishing entry equal to one.} The desired cocycle $u_{1,2}(t)$ is 
determined by $V_t$ as 
$$
\begin{bmatrix} 
	0 & 0  \\
	u_{1,2}(t) & 0 \\
\end{bmatrix}
$$
or in matrix notation 
\begin{equation}
u_{1,2}(t)\otimes e_{2,1}=\sigma_{\Phi}^t(1\otimes e_{2,1})
\end{equation}
In words, what we do in order to compare two states $\phi_1$ and $\phi_2$ is to define, in the bigger algebra $A\otimes M_2(\mathbb{C})$, two "copies" of $A$ using the projections $p$ and $q$ i.e. using $e_{1,1}$ and $e_{2,2}$ with the state $\phi_1$ acting on one copy and the state $\phi_2$ acting on the other. The cocycle relating both states is determined by $\sigma_{\Phi}^t(1\otimes e_{2,1})$ with $\Phi=\frac{1}{2}\sum\phi_i$ a state defined on the extended algebra $A\otimes M_2(\mathbb{C})$. In this example we easily get the relation
\begin{equation}\label{cocyclerelation}
V^*\sigma_{\Phi}^t(V)=(u_{1,2}^t \otimes e_{1,1})
\end{equation}
for $V=(1\otimes e_{2,1})$. 

The crucial result of Connes's cocycle is to prove that the modular automorphism $\sigma_{\phi}^t$ is {\it unique} up to inner automorphisms. This means that the dependence on the particular background $\phi$ reduces to the action of the inner automorphism defined by the cocycle $t\rightarrow u_t$. However the cocycle defined by the map $t\rightarrow u_t$ is {\it not a unitary operator} in $A\otimes M_2(\mathbb{C})$. But recall that in order to prove background independence we need to prove the {\it unitary equivalence} of different backgrounds.

As we will show to prove background independence will require to extend the algebra as $A\otimes F_{\infty}$ and to associate with two arbitrary backgrounds $\phi_1$ and $\phi_2$ a {\it unitary} operator $V_{1,2}$ in $A\otimes F_{\infty}$. It is in order to do that that weights will become important.

\subsection{A hint for background independence}
In \cite{Connes1} we find a theorem\footnote{Lemma 1.2.5 in \cite{Connes1}.} that provides the key hint for background independence. 
Here we will simply recall the theorem leaving for the next section the underlying physics meaning. The theorem says that given a continuous map $t\rightarrow u_t$ satisfying (\ref{cocycle1}) for some state $\phi_1$ i.e. such that $u_{t+t'}= u_t \sigma_{\phi_1}^t(u_t')$ there exist:

 i) another state $\phi_2$ such that $\sigma_2^t=u_t\sigma_1^tu_t^*$ and 
 
 ii) a unitary $V$ in $A\otimes F_{\infty}$ and a weight $\omega$ on $F_{\infty}$ such that
\begin{equation}\label{hint}
(u_t\otimes 1)= V\sigma_{\bar \omega}^t (V^{*})
\end{equation}
for $\bar\omega=\phi_1\otimes\omega$.

Note that relation (\ref{hint}) is very similar to relation (\ref{cocyclerelation}). However there are important differences. The first one is that we use the type $I_{\infty}$ factor $F_{\infty}$ instead of the much simpler {\it two q-bits} algebra $M_2(\mathbb{C})$. The second difference is that we use the {\it weight} $\bar\omega$ instead of the {\it state} $\Phi$ used in (\ref{hint}). 

Note that if the weight $\bar\omega$ in (\ref{hint}) is {\it unique} (up to unitary equivalence) what (\ref{hint}) is telling us is that by extending the algebra $A$ into $A\otimes F_{\infty}$ the cocycle $u_{1,2}(t)$ relating two arbitrary states on $A$ can be represented by the {\it unitary} element $V_{1,2}$ in $A\otimes F_{\infty}$. Indeed it is easy to see that if the weight $\bar\omega(\phi)=\phi\otimes \omega$ satisfies (\ref{hint}) for a generic state $\phi$ then the weights $\bar\omega(\phi_1)$ and $\bar\omega(\phi_2)$ for two different states $\phi_1$ and $\phi_2$ are equivalent and in that sense the weight $\bar\omega$ satisfying (\ref{hint}) is unique. What is a harder is to discover what is the weight $\omega$ on $F_{\infty}$ such that $\bar\omega=\phi\otimes \omega$ satisfies (\ref{hint}).

Thus if we identify the weight $\omega$ on $F_{\infty}$ such that for any $\phi$ on $A$ the weight $\bar\omega=\phi\otimes \omega$ satisfies (\ref{hint}) then we will have identified on $A\otimes F_{\infty}$ a {\it background independent weight}. Physically this means that {\it background independence} relative to states $\phi$ on $A$ requires:

i) To {\it add} the extra $I_{\infty}$ factor $F_{\infty}$, which we can read as the addition of some extra degree of freedom and

ii) To use a particular weight $\omega$ on $F_{\infty}$ that we can interpret as defining a particular dynamics on the extra degree of freedom.

The reader familiar with the approach of \cite{Witten0} to de Sitter will easily identify the extra degree of freedom associated to $F_{\infty}$ as somehow representing an external observer. We will discuss this interpretation later. For the time being we will think in this $F_{\infty}$ factor as needed to achieve {\it background independence} with respect to arbitrary changes of the state $\phi$ on $A$ \footnote{In a certain sense when we deal with {\it properly infinite} algebras the {\it observer}, as being represented by $F_{\infty}$, can be abstractly thought as existing by default. Indeed we can always represent the properly infinite algebra $A$ as $A= B\otimes F_{\infty}$ for some $B$ properly infinite and  {\it isomorphic} to $A$.}.

In order to identify the desired weight $\omega$ and its physical meaning we need to go through some preliminary technicalities.

{\bf Remark} Note that the state $\Phi$ on $A\otimes M_2(\mathbb{C})$ defined in (\ref{state}) can be interpreted as representing a {\it quantum superposition} of the two backgrounds $\phi_1$ and $\phi_2$. This state is now replaced by the {\it unique} weight $\bar\omega$. Hence $\bar\omega$ appears, already at this level of the discussion, as a good candidate to define a potential quantum superposition of all the backgrounds for the type $III$ algebra $A$.

\subsection{Dominant weights}
\subsubsection{The $F_{\infty}$ factor and the CCR}
Let us start discussing the basic properties of the added factor $F_{\infty}$ and its natural interpretation as representing an added quantum degree of freedom.
The type $I_{\infty}$ factor $F_{\infty}$ is defined as the algebra of bounded operators acting on the Hilbert space $L^2(\mathbb{R})$. Quantum mechanically we can think of $L^2(\mathbb{R})$ as the space of quantum states of some elementary system with one degree of freedom with {\it coordinate} $t\in {\mathbb{R}}$ and with functions $f(t)\in L^2(\mathbb{R})$ as Schrodinger wave functions. Let us now define for any $s\in {\mathbb{R}}$ the following {\it unitary} operators in $F_{\infty}$:
\begin{equation}\label{momentum}
(U_sf)(t)=f(s+t)
\end{equation}
and
\begin{equation}\label{position}
(V_s f)(t)= e^{ist} f(t)
\end{equation}
In standard quantum mechanics you will define $U_s=e^{is p}$ for $p$ representing the {\it momentum} operator and $V_s=e^{is\hat{t}}$ for $\hat t$ representing the {\it position} operator i.e $(\hat t f)(t) = t f(t)$. The corresponding Heisenberg algebra is given by $[p,\hat t]= -i$\footnote{In what follows we will use the convention $\hbar=1$. See comment in section7.5.} and $U$ and $V$ define the standard Weyl representation of the canonical commutation relations (CCR) in terms of bounded operators.

On $F_{\infty}$ we can define {\it weights}. As already discusses a weight is defined by a linear map $\omega$ on $F_{\infty}^{+}$ with values in $[0,+\infty]$ satisfying $\omega(x+y)=\omega(x)+\omega(y)$ for $x$ and $y$ in $F_{\infty}^{+}$ and $\omega(\lambda x)=\lambda \omega(x)$ for $\lambda > 0$ and $x \in F_{\infty}^{+}$. Intuitively a weight extends the notion of state to the standard trace on $F_{\infty}$ that can take values in $[0,\infty]$.

For each weight $\omega$ on $F_{\infty}$ we can define the corresponding modular automorphism $\sigma_{\omega}^t$ ( see \cite{Connes1}). This can be also done for the weight defined by the standard trace that we can denote $\sigma_{Tr}^t$. Now we can define Connes cocycle relating both weights i.e. the unitary $u_t$ such that $\sigma_{\omega}^t=u_t\sigma_{Tr}^t u_t^{*}$. We will now define the {\it particular weight} $\omega$ on $F_{\infty}$ by requiring
\begin{equation}\label{condition}
u_t=U_t
\end{equation}
for $U_t$ defined in (\ref{momentum}) and for $u_t$ the cocycle relating $\omega$ and the weight defined by the standard trace on $F_{\infty}$.

The condition (\ref{condition}) leads to
\begin{equation}\label{weight2}
\omega(x)= Tr(\rho x)
\end{equation}
for $x\in F_{\infty}^+$ i.e. $x\geq0$ and with $\rho^{it}=U_t$. That implies $\rho=e^{p}$ with $p=-i\frac{d}{dt}$. Note that (\ref{weight2}) has a nice intuitive meaning. Indeed, if we insist in defining a formal Hamiltonian associated to this elementary quantum mechanical system as $H_r=i\frac{d}{dt}$ then we get
\begin{equation}\label{observerweight}
\omega(x)=Tr(e^{-H_r} x)
\end{equation}
that looks like the canonical Gibbs definition of thermal expectation values for elements in the "added" algebra $F_{\infty}$ \footnote{Notice that if we assign to $t$ dimensions of length we can define the dimensionless generator $\rho$ in (\ref{weight2}) as $\beta H_r$. In the notation of \cite{Witten1} the role of $H_r$ is replaced by the "observer" hamiltonian $m+q$. See discussion in 7.6.1 of the present note.}. 

It is easy to check that the {\it scale} transformation $\omega\rightarrow \lambda\omega$ for $\lambda > 0$ is implemented by the, canonically conjugated, unitary transformation $V_s$ for $s=-\ln \lambda$ as \footnote{Note that we can interpret $V_s$ in terms of $\mathbb{R}_{+}^{*}$ with the dual pairing $<s,t>=e^{ist}$.}
\begin{equation}
V_s(\omega)=\lambda\omega
\end{equation}
where we define the weight $V_s(\omega)$ by $V_s(\omega)(x)=\omega(V_s x V_s^{*})$.
\subsubsection{The dominant weights}

The weight $\omega$ on $F_{\infty}$ defined in (\ref{observerweight}) is the weight we were looking for, namely the one for which the weight $\bar\omega=\phi\otimes \omega$ on $A\otimes F_{\infty}$ for $\phi$ a generic faithful and normal state on $A$ satisfies (\ref{hint}). We already know why this weight is interesting. Namely it is unique up to unitary equivalence.  By that we mean that for two states $\phi_1$ and $\phi_2$ on $A$ the weights
$\bar \omega_1=\phi_1\otimes \omega$ and $\bar\omega_2=\phi_2\otimes \omega$ are related by a unitary transformation \footnote{Or in more precise words are in the same equivalence class of weights (see \cite{Connes2} and discussion in the next paragraph).}. 

In summary we observe that the weight $\omega$ on $F_{\infty}$ needed to define the {\it background independent} weight $\bar\omega$ is the formal analog of the Gibbs canonical distribution $Tr(x\rho)$ for $\rho^{it}=U_t$ with $U_t$ the CCR Weyl bounded operator associated to the momentum.

In \cite{Connes2}
an equivalence relation for weights is introduced by analogy with the equivalence relation between projections. Thus for two different weights $\omega_1$ and $\omega_2$ of a properly infinity von Neumann algebra $A$ \footnote{Recall that by that we mean that the identity of $A$ is properly infinity i.e. $A$ is of type $III$.} we will say that both weights are {\it equivalent} represented by $\sim$ if there exist a partial isometry $u$ \footnote{Recall that for a partial isometry $u$ both $u^*u$ and $uu^*$ are projections. Thus two projections $p$ and $q$ are equivalent if there exist a partial isometry $u$ such that $u^*u=p$ and $uu^*=q$.}  such that $\omega_1(x)=\omega_2(uxu^*)$. For a given weight $\omega$ we will denote $[\omega]$ the corresponding {\it equivalence class}\footnote{The set of equivalence classes define a Boolean algebra with $[\alpha]+[\beta]=[\alpha+\beta]$ provided the support of $\alpha$ and the support of $\beta$ are orthogonal.}. Equivalently two weights $\omega_1$ and $\omega_2$ are equivalent if there exist a partial isometry $u$ such that $u^*u$ is equal to the support of $\omega_1$ and $uu^*$ is equal to the support of $\omega_2$.

Now we can define the flow of weights associated to $A$ as the "scale transformation" $F_{{\cal{A}}}(\lambda)$ on the equivalence classes of weights
\begin{equation}\label{flow1}
F_{{\cal{A}}}(\lambda): [\omega] \rightarrow [\lambda \omega]
\end{equation}
for $\lambda \geq 0$ and $\omega$ a weight of $A$ \footnote{We will discuss a more precise definition of the flow of weights in next section.}.

Let us now define the {\it dominant weight} $\bar \omega$ on the algebra of interest namely, $A\otimes F_{\infty}$ for $A$ the type $III_1$ factor associated to the de Sitter static patch. As said this is the weight satisfying (\ref{hint}), namely
\begin{equation}
\bar\omega = \phi\otimes \omega
\end{equation}
with $\omega$ the weight defined in (\ref{observerweight}) as solving the condition (\ref{condition}). For the equivalence relation defined above we observe that for the dominant weight $\bar\omega\sim \lambda \bar\omega$.

We know that the dominant weight $\bar\omega$ defined above is interesting because is independent ( up to unitary equivalence) of the state $\phi$ on $A$ and in that sense is unique. In addition
we easily observe that this weight transforms under the unitary $(1\otimes V_{s})$ for $V_{s}$ defined in (\ref{position}) as
\begin{equation}
\bar \omega \rightarrow e^{-s}\bar\omega
\end{equation}
This means, as already said, that $\bar\omega$ and $\lambda \bar\omega$ are in the same equivalence class $[\bar\omega]$\footnote{This scale property characterizes uniquely the weight $\bar\omega$ ( see theorem 1.1 in \cite{Connes2}).}. 

In essence what we have achieved with the weight $\bar\omega$ on $A\otimes F_{\infty}$ is, as already stressed, {\it background independence} in the sense of independence (up to unitary equivalence) on the state $\phi$ on $A$. Let us keep in mind that in order to achieve this independence on the state $\phi$ we have paid a double price, namely to extend the algebra with the factor $F_{\infty}$ and to work with weights which in essence means that we are using the standard, and not finite, trace on $F_{\infty}$.

Before ending this section let us recall few properties of $\bar\omega$ that will be important in the future discussion.

\subsubsection{Intregrability}
By an integrable weight on a type $III$ algebra $A$ we mean a weight $\varphi$ such that
the set of elements $Q_{\varphi}$ in $A$ for which
\begin{equation}
\int_{-\infty}^{+\infty}\sigma_{\varphi}^t(x^*x)dt
\end{equation}
exists is dense in $A$ \footnote{For a more rigorous definition see \cite{Connes2} II-2.}. Note that for an integrable weight the operator
\begin{equation}\label{conditional}
E_{\varphi}(x)= \int_{-\infty}^{+\infty} \sigma_{\varphi}(x)dt
\end{equation}
makes sense for any $x\in Q_{\varphi}$. Moreover $\sigma_{\varphi}^t(E_{\varphi}(x))= E_{\varphi}(x)$ i.e. $E_{\varphi}(x)$ is invariant under the action of the modular automorphism $\sigma_{\varphi}^t$. The set of elements invariant under $\sigma_{\varphi}$ is known as the {\it centralizer} of $\varphi$.

\subsubsection{The centralizer}
The centralizer of $\bar\omega$ is the set of elements $x$ in $A\otimes F_{\infty}$ such that
$\sigma_{\bar\omega}^t(x)=x$. This defines a subalgebra of $A\otimes F_{\infty}$ that we will denote ${\cal{C}}_{\bar\omega}$. This algebra is acting on $L^2(\mathbb{R},H_{\phi})$ and it is generated by $(1\otimes U_t)$ for $U_t$ defined in (\ref{momentum}) and by the maps $\hat a_{\phi}: t\rightarrow \sigma_{\phi}^t(a)$. 

For two different weights $\phi_1$ and $\phi_2$ we can define a unitary transformation $V_{1,2}: L^2(\mathbb{R},H_{\phi_1})\rightarrow L^2(\mathbb{R},H_{\phi_2})$ using the corresponding cocycle $u_{1,2}(t)$. Namely we can define the unitary operator $V^{1,2}$
in terms of the cocycle $u_{1,2}(t): H_{\phi_1}\rightarrow H_{\phi_2}$ as
\begin{equation}\label{unitaria}
(V_{1,2} f_1)(t) = u_{1,2}(t)( f_1(t))
\end{equation}
with $f_1(t)$ in $L^2(\mathbb{R},H_{\phi_1})$.

Now it is easy to check that conjugation of $\hat a_{1}$ by $V_{1,2}$ produces $\hat a_{2}$. This allows us to identify the centralizer ${\cal{C}}_{\bar\omega}$ with the crossed product
\begin{equation}\label{cross}
{\cal{C}}_{\bar\omega}= A \rtimes_{\sigma_{\phi}}{\mathbb{R}}
\end{equation}

Finally, in order to justify (\ref{cross}) let us recall that the crossed product $A\rtimes_{\sigma_{\phi}} {\mathbb{R}}$ of $A$ by the modular automorphism $\sigma_{\phi}^t$ is generated by $(1\otimes U_t)$ with $U_t$ defined in (\ref{momentum}) and by operators $\hat a$ in $A\otimes F_{\infty}$ associated to any operator $a\in A$ and acting on $L^2(\mathbb{R},H_{\phi})$ as
\begin{equation}\label{dressed}
(\hat a f)(t)=\sigma_{\phi}^t(a) f(t)
\end{equation}
for any $f\in L^2(\mathbb{R},H_{\phi})$.

{\bf Remark} Note that the dominant weight $\bar\omega$ on the centralizer satisfies the KMS condition ( see footnote 25) for formal thermal correlators at infinite $\beta=0$ temperature. However the definition of $\bar\omega$ contains a reference to an arbitrary finite temperature $\beta$ that is the one defining the $F_{\infty}$ part $\omega$ of $\bar\omega$ ( see footnote 35). Note that this $\beta$ cannot be made equal to zero without destroying the nature of $\bar\omega$ as a dominant weight i.e. as defining the background independent weight.
\subsubsection{The flow of weights and the role of the CCR}
Let us note that the centralizer ${\cal{C}}_{\bar\omega}=A\rtimes_{\sigma_{\phi}}{\mathbb{R}}$ is not anymore a factor. Indeed the {\it center} ${\cal{Z}}_{\bar\omega}$ of ${\cal{C}}_{\bar\omega}$ is the whole algebra generated by the $U_t$. However in the full algebra $A\otimes F_{\infty}$ we have in addition to $U_t$ the canonical conjugated $V_s$ satisfying the  Weyl CCR relation
\begin{equation}
V_sU_t=e^{ist}U_tV_s
\end{equation}
What is the role of $V_s$ ? As discussed it
allows us to define an {\it automorphism} of ${\cal{C}}_{\bar\omega}$ as
\begin{equation}\label{auto}
\theta_s: x\rightarrow (1\otimes V_s)x(1\otimes V_s)^*
\end{equation}
On the center ${\cal{Z}}_{\bar\omega}$ it reduces to simply $\theta_s: U_t\rightarrow e^{ist}U_t$. This obviously induces the flow of weights $\theta_s: \bar\omega\rightarrow \lambda\bar\omega$ with $\lambda=-\ln s$. Thus in our case the flow of weights reduces to the action of $\theta_s$ on the center ${\cal{Z}}_{\bar\omega}$ of the centralizer ${\cal{C}}_{\bar\omega}$. A consequence of the former discussion is the dual crossed product representation\footnote{This is a consequence of Takesaki duality, \cite{Takesaki} and \cite{Witten1} appendix B.}
\begin{equation}\label{dual}
A\otimes F_{\infty}= {\cal{C}}_{\bar\omega}\rtimes_{\theta_s}{\mathbb{R}}
\end{equation}

\section{The Trace the  Dressing and Affiliations}
Summarizing what we have achieved in the previous section is to associate with the type $III_1$ factor $A$ describing the de Sitter static patch the {\it background independent subalgebra} ${\cal{C}}_{\bar\omega}$ for $\bar\omega$ the {\it unique} dominant weight on $A\otimes F_{\infty}$. The KMS property for $\bar\omega$ implies that $\bar\omega$ reduced to the centralizer ${\cal{C}}_{\bar\omega}$ defines a semifinite trace $\tau$ defined as
\begin{equation}\label{deftrace}
\tau(x)=\bar\omega(x)
\end{equation}
for $x\in {\cal{C}}_{\bar\omega}$. Thus we have obtained, using background independence: 
\begin{itemize}
\item A background independent algebra that will define the set of background independent physical observables and 
\item A semi finite trace on this algebra on the basis of which we can define some entropic notions.
\end{itemize}

As already discussed the automorphism $\theta_s$ defined in (\ref{auto}) scale the trace as
\begin{equation}
\tau(\theta_s(x))=e^{-s}\tau(x)
\end{equation}
The algebra ${\cal{C}}_{\bar\omega}$ is generated by the $U_t$ in the center and the elements $\hat a$ defined in (\ref{dressed}). Since $\hat a$ acting on $L^2(\mathbb{R},H_{\phi})$ is defined by $(\hat a f)(t) = \sigma_{\phi}^t(a) f(t)$ we can define the {\it dressing map} $d_{\phi}: A \rightarrow A\otimes F_{\infty}$ as
\begin{equation}\label{dressing}
d_{\phi}: a\rightarrow \sigma_{\phi}^{\hat t} (a)
\end{equation}
where we define $\sigma_{\phi}^{\hat t}(a)=e^{ih_{\phi}\hat t}ae^{-ih_{\phi}\hat t}$ for $\hat t$ formally defined as $\hat t f(t)=tf(t)$ for $f(t)\in L^2(\mathbb{R})$. Note that this {\it dressing} is {\it background independent} in the sense that
\begin{equation}
d_{\phi_1}(a) =V_{1,2}d_{\phi_2}(a)V_{1,2}^*
\end{equation}
for $V$ the unitary transformation in $A\otimes F_{\infty}$ defined in (\ref{unitaria}).

Recall from the previous section that due to Takesaki duality we have a different representation of $A\otimes F_{\infty}$ as the crossed product ${\cal{C}}_{\infty}\rtimes_{\theta_s}{\mathbb{R}}$ and that $E$ defined in (\ref{conditional}) for $\bar\omega$ defines a {\it conditional expectation} from $A\otimes F_{\infty}$ into ${\cal{C}}_{\bar\omega}$. We can think of this projection as defining a different dressing operator $D$ associated to the dominant weight $\bar\omega$. Again this dressing $D$ is background independent due to the background independence of $\bar\omega$ \footnote{Using an analogy with gauge theories we can think of $D$ as playing a similar role to the projection in the gauge bundle ${\cal{A}}\rightarrow \frac{\cal{A}}{G}\rightarrow G$ \cite{AJ} with ${\cal{A}}$ the space of gauge connections and $\frac{\cal{A}}{G}$ the space of gauge orbits.}.

Finally let us introduce the notion of {\it affiliations}. For any weight $\varphi$ on ${\cal{C}}_{\bar\omega}$ we can define the affiliated self adjoint operator $h_{\varphi}$ by the relation
\begin{equation}\label{affiliation}
\tau(h_{\varphi} \cdot) = \varphi(\cdot)
\end{equation}
In physics terms and when we use for $\tau$ a {\it finite trace} then the relation (\ref{affiliation}) defines the {\it density matrix} $h_{\varphi}$ associated to the state $\varphi$. In our present case $\tau$ is semi finite and consequently $h_{\varphi}$ is not a standard density matrix \footnote{As we will briefly discuss in section 7.4 the definition of the standard von Neumann entropy in terms of the semifinite trace $\tau$ and the affiliated operators can only be defined in those cases where $\ln h_{\varphi}$ is in ${\cal{C}}_{\bar\omega}^{+}$ where $\tau$ is well defined as a weight.}

One important technical point about which we should be very careful is not to identify semifinite and normal weights with vectors states. Obviously for any $\xi\in{\cal{H}}$ where we denote ${\cal{H}}$ the Hilbert space $L^2(\mathbb{R},H_{\phi})$ we can define a linear map $\varphi_{\xi}$ as
\begin{equation}\label{repreones}
\varphi_{\xi} (a)= \langle a\xi,\xi\rangle
\end{equation}
with the scalar product the one defining ${\cal{H}}$. The so defined map is finite. In particular if $\xi$ is an honest vector state in ${\cal{H}}$ of finite norm the value of so defined map $\varphi_{\xi}(I)$ on the identity is finite. For the semifinite weights we are using this is clearly not the case thus we cannot use (\ref{repreones}) to represent the semifinite weights for instance $\bar\omega$. Thus these weights are {\it not} associated to particular vector states in ${\cal{H}}$. However we can get, in general, a vector state representation of the  semi-finite weight of the type (\label{repreones}) if we work {\it not in ${\cal{H}}$ but in the extended Hilbert space of square summable sequences $l^2({\mathbb{N}},{\cal{H}})$}.

{\bf Remark} Note that for the type $II_{\infty}$ description of the eternal black hole in \cite{many5} the semi-finite weight defining the semi-finite type $II_{\infty}$ trace cannot be represented as a vector state in ${\cal{H}}$ although it can be represented as a square summable sequence in $l^2({\mathbb{N}},{\cal{H}})$ \footnote{This is likely related to the microcanonical version used in \cite{many5}.}. 

\subsection{Marginal digression: Some general remarks on dressing and global charges}
As a small independent digression let us briefly discuss the meaning of dressing for global symmetries. Let us imagine a global symmetry group $G$ acting on a von Neumann algebra $A$ i.e. an automorphism $\alpha_g$ of $A$ with $g\in G$ and let us assume $A$ is acting on a Hilbert space $H$. The invariant subalgebra is $A^{G}$. Moreover if the state $\phi$ on $A$ used to define the expectation values is G-invariant the action of the corresponding modular automorphism $\sigma_{\phi}^t$ is leaving $A^{G}$ invariant i.e. $\sigma_{\phi}^t (A^{G}) = A^G$. This means that the {\it modular Hamiltonian} commutes with the charge generating the global symmetry $G$.

If we consider now the crossed product for the action of $G$ on $A$
\begin{equation}
A \rtimes_{\alpha_{g}} {G}
\end{equation}
as an algebra acting on $L^2(G,H)$ \footnote{Defined relative to the Haar measure on $G$.} we can define the corresponding {\it dressed} operators $d_{G}(a)$ using the analog of (\ref{dressing}) namely $(d_{G}(a) f)(g)=\alpha_{-g}(f(g))$ for $g\in G$ and $f(g)\in L^2(G,H)$ . Let us denote $\hat A$ the so defined algebra i.e. the image $d_{G}(A)$ in $A \rtimes_{\alpha_{g}} {G}$. 

In \cite{many6.1} we have suggested to replace $A^G$ by the corresponding dressed algebra $\hat A$ for the dressing defined by $d_{G}$. In particular we suggested to interpret Aharonov-Susskind (AS) \cite{AS} approach to superselection charges as the replacement
\begin{equation}
AS: A^G \rightarrow \hat A
\end{equation}
Note that while using $A^G$ as the set of physical observables makes superposition of states with different charges unobservable this is not the case when se use the dressed ( defined using the crossed product) algebra $\hat A$. Note that this dressing requires the addition of the quantum reference frame \cite{RF} represented by the extra degree of freedom, with Hilbert space $L^2(G)$ used in the definition of the crossed product.

\section{The OPE-algebra and the GNS-representation}

After this rather abstract discussion is time to come back to some basic questions that we have ignored until now. In particular how a generic observer is defining the $C^*$ algebra $A$ we have used in our discussion and what, if any, is the physics meaning of the extension of this algebra into $A\otimes F_{\infty}$. 

Our starting point was to identify $A$ with the $C^*$ algebra of observables that a generic external and ideal observer can locally measure. This is a rather vague definition before identifying what qualifies as an observer. Following many authors ( see \cite{Witten0} for an insightful discussion ) we can think of an observer as a timelike trajectory $x(\tau)$ in some  classical space time background that uses her proper time $\tau$ to parametrize local observations ( measurements) performed in her neighborhood. Let us denote these observables as $O(\tau)$. The observer can easily provide the set of $O(\tau)$ with the structure of a vector space, however it is less easy to identify how the observer equip the set of local observations $O(\tau)$ with the structure of an algebra. To define this algebra the observer can use some QFT operator product expansion OPE at least in the short time limit. The leading term of the product $O(\tau)O(0)$ in the short time limit, assuming you have defined a $(*)$ adjoint operation and that $O(\tau)$ is self adjoint, should be some function $C(\tau)1$ for $1$ representing the identity. If in the short time limit you assume conformal invariance then $C(\tau)\sim \frac{1}{\tau^{2\Delta_{O}}}$ for some $\Delta_{O}$. In any case this QFT OPE algebra leads to the familiar UV divergences in QFT. So, how to define a "regular" OPE and the corresponding $C^*$ algebra? 

The natural recipe in QFT is to smear (in time) the observables $O(\tau)$ \footnote{As discussed in \cite{Witten-o} this smear in time leads to a good OPE i.e. a finite product of operators. Note that this is not the case if we just smear in space.}, in the sense of distributions, with some test functions $f(\tau)$. Note that before going into any representation of the algebra in terms of bounded operators in a Hilbert space we need to use test functions to integrate the short distance divergences predicted by the OPE. This means that the OPE multiplication rule is defined not for the $O(\tau)$ but for some functions $\tau\rightarrow O(\tau)$ satisfying the OPE integrability properties i.e. defining a good and finite OPE. We can think the test functions as defining particular functions $\tau\rightarrow O(\tau)$ square integrable with respect to the norm of the $C^*$ algebra. What algebra are these functions defining ? A natural answer is that they define the crossed product $A\rtimes_{\sigma_{\phi}}{\mathbb{R}}$ or equivalently that they represent dressed observables. As discussed the key of this identification is to be {\it background independent} in the sense of being ( up to unitary equivalence) independent of what state $\phi$ we use.

But as discussed the so defined crossed product contains, in addition to the "dressed operators" associated to functions $\tau\rightarrow O(\tau)$, the subalgebra generated by the $U_{\tau}$ acting on $F_{\infty}$ and generating translations of the proper time coordinate $\tau$. In that sense they can be interpreted as defining some formal clock Hamiltonian provided we identify the coordinate $\tau$ with a proper time.

In order to complete the picture we need to define expectation values for the dressed operators as well as for the $U_{\tau}$ subalgebra and we want to do it preserving background independence. This requires to define a weight on $A\otimes F_{\infty}$ that transforms unitarily under changes of background. We know that this weight is unique and it is the dominant weight $\bar\omega$ discussed in previous sections. 

At first sight it looks that we have finished our job. Namely we have identified a good OPE algebra of dressed operators and a unique weight defining expectation values that is background independent; the dominant weight. However, and indeed fortunately, this is not the end of the story.

Once we have added the factor $F_{\infty}$ we have being forced to work with weights. The reason is simply the lack of a finite trace on $F_{\infty}$. But weights lead to "divergences" in the sense of semi finite traces  valued in $[0,+\infty]$. Mathematically this is expressed saying that the crossed product algebra we have defined is a type $II_{\infty}$ algebra. So we need to face the question: How to tame these type $II_{\infty}$ "divergences" ?

The answer suggested in \cite{Witten1} is to define a projector $\Pi$ on $L^2(\mathbb{R})$ such that $\Pi(e^{-ip\tau})=0$ for $p<0$ and $\Pi(e^{-ip\tau})=e^{-ip\tau}$ for $p>0$ \footnote{The integral representation of this projector is given in \cite{Witten0}.} and to define a type $II_1$ algebra as $\Pi(A\rtimes_{\sigma{\phi}}{\mathbb{R}})\Pi$.

Although the use of $\Pi$ makes a lot of physics sense and can be interpreted as defining a {\it positive} clock Hamiltonian i.e. with the generator of $U_{\tau}$ having a positive spectrum, the construction is more general. Indeed we could use any finite projection $P$ in $A\rtimes_{\sigma{\phi}}{\mathbb{R}}$ where by that we mean any projection with finite type $II_{\infty}$ trace and to define the type $II_1$ algebra as $P(A\rtimes_{\sigma{\phi}}{\mathbb{R}})P$.

Now the background independent $\bar\omega$ defines for any finite projection $P$ a type $II_1$ trace on $P(A\rtimes_{\sigma{\phi}}{\mathbb{R}})P$. It is at this point that we can try to define {\it quantum gravity} effects using $\bar\omega$ to {\it compare} different and inequivalent finite projections in the way we used in section 3.1 to define Connes cocycle. The key observation is that the {\it background independent} $\bar\omega$ allows us to compare different type $II_1$ algebras that are not {\it unitarily equivalent}. We will use this observation in section 7 to define a toy model of quantum gravity for an ensemble of de Sitter Universes. Before that we will briefly pause to discuss the quantum notion of {\it clocks}.

\section{Interlude : Clocks in quantum mechanics}
Intuitively by a quantum mechanical clock we mean a system ( potentially elementary and therefore with Hilbert space $L^2(\mathbb{R})$ ) characterized by 

i) a position operator observable, let us say $q$, that we use to tell the time 

ii) a clock Hamiltonian $H_{clock}$ that defines the {\it Tic-Tac} of the clock by the Heisenberg equation $\dot q=[q,H_{clock}]$ and 

iii) a self adjoint {\it time operator} $\hat t_{clock}$ depending on $q$ and such that on a reduced subspace ${\cal{S}}_{clock}$ of $L^2(\mathbb{R})$ satisfies $\langle \psi| [\hat t_{clock},H_{clock}]|\psi\rangle = -i\hbar$ for $|\psi\rangle \in {\cal{S}}_{clock}$. 

The reason to impose condition iii) is to give an observational meaning to the energy-time uncertainty relation $\Delta(t)\Delta(E)\geq \hbar$ for $t$ the eigenvalue of $\hat t_{clock}$ on ${\cal{S}}_{clock}$. The obvious reason, forcing us to introduce ${\cal{S}}_{clock}$, is that the condition $[\hat t_{clock},H_{clock}]=i\hbar$ implies, for self adjoint operators, that the spectrum of $H_{clock}$ is the full real line and consequently it cannot be a positive physical Hamiltonian.

An early attempt to define $\hat t _{clock}$ and ${\cal{S}}_{clock}$ is due to Aharonov and Bohm \cite{AB2} in the spirit of the Mandelstam-Tamm \cite{MT} approach to the energy-time uncertainty relation. In this section we will approach the so defined {\it clock problem} in the algebraic framework.

\subsection{Clocks and Weights}
In order to define a {\it clock} let us consider the factor $F_{\infty}$ used in previous sections but equipped with a self adjoint and positive Hamiltonian $H_{clock}$. 
Let us now try to define a weight $\omega_{clock}$ formally associated to the physical i.e. positive, clock Hamiltonians $H_{clock}$. We could formally define the weight   $\omega_{clock}$ on $F_{\infty}$ as 
\begin{equation}
\omega_{clock}(x)=Tr(xe^{H_{clock}})
\end{equation}
Superficially the so defined clock-weight only differs from the weight $\omega$ in that now we represent the translations by $t$ in $L^2({\mathbb{R}})$ using the positive Hamiltonian $H_{clock}$. Can we achieve background independence using a $\bar\omega_{clock}$ defined as $\phi\otimes \omega_{clock}$?

The first thing to be noticed is that the weight $\bar\omega_{clock}$ is not satisfying (\ref{hint}). Indeed
when we decide to use a physical {\it positive} clock Hamiltonian $H_{clock}$ to define the weight $\omega_{clock}$ on $F_{\infty}$ the cocycle relating $\sigma_{\omega^{clock}}^t$ and $\sigma_{Tr}^t$ will be determined by $H_{clock}$ but this is only representing translations by $t$ on the {\it reduced space}, let us call it ${\cal{S}}_{clock}(\mathbb{R})$, defining the solutions of the associated Schrodinger equation
\begin{equation}\label{schrodinger}
\frac{df}{dt}=H^{clock}f(t)
\end{equation}
and therefore $\bar\omega_{clock}$ will not be a dominant weight. From a physics point of view the weight $\omega_{clock}$ will have as support only those bounded operators acting on the observer Hilbert space that are closed in the space of solutions to the Schrodinger equation (\ref{schrodinger}). Thus the support of $\bar\omega_{clock}$ is contained in the support of the dominant weight $\bar\omega$. The best we can do to keep background independence using a physical positive Hamiltonian $H_{clock}$ is to reduce the Hilbert space $L^2(\mathbb{R},H_{\phi})$ to ${\cal{S}}_{clock}(\mathbb{R}) \otimes H_{\phi}$ and to assume that it exists a projection $P_{clock}$ defining ${\cal{S}}_{clock}(\mathbb{R}) \otimes H_{\phi}$.

At this point it is instructive to understand  the meaning of the formal operator $\hat t_{clock}$ introduced in \cite{AB2} satisfying $[\hat t_{clock},H_{clock}]=i\hbar$ {\it but} on a reduced set of clock states ${\cal{S}}_{clock}$. Recall from \cite{AB2} that for the simplest example corresponding to a free clock Hamiltonan $H_{clock}=\frac{1}{2M} p^2$ with position operator $q$ and with a mass scale $M$ roughly representing the clock mass, we can write $\hat t_{clock}=\frac{1}{2M}(q\frac{1}{p}+\frac{1}{p}q)$. This $\hat t_{clock}$ is not defined on those states in $L^2(\mathbb{R})$ such that $\langle M\dot q \rangle =0$. Physically we can think of $\hat t_{clock}$ as $\frac{q}{\langle \dot q \rangle}$ with $\dot q=[H_{clock},q]$. In this case we can define ${\cal{S}}_{clock}$ as those wave functions with a fixed value of $\langle \dot q \rangle$ in a small interval determined by the variance of $\dot q$ that we will assume very small for large enough $M$. Hence in this approximation we will set ${\cal{S}}_{clock}$ by the value of $\langle \dot q\rangle$ and we will denote this space as ${\cal{S}}_{clock}(\langle \dot q\rangle)$. 

Let us now assume that there exist a finite projection $P_{\langle \dot q\rangle}$ from $L^2(\mathbb{R})$ into ${\cal{S}}_{clock}(\langle \dot q\rangle)$ \footnote{Althouh this assumption is from clear to the author it can clarify the logic of the discussion.}. In this case we can define the {\it clock centralizer} ${\cal{C}}_{clock}$ as $P_{\langle \dot q\rangle} {\cal{C}}_{\bar\omega}P_{\langle \dot q\rangle}$. In such a case this clock centralizer could be a type $II_1$ algebra if the projection $P_{\langle \dot q\rangle}$ is in ${\cal{C}}_{\bar\omega}$ and it is finite relative to the trace $\tau$ defined in (\ref{deftrace}). 

{\bf Remark} It is important to notice that any introduction of a formally self adjoint operator $\hat t_{clock}$ changes the notion of dressing. In particular using (\ref{dressing}) we can define a {\it clock dressing} as
\begin{equation}
d_{clock}(a) = e^{i\hat t_{clock} h_{\phi}} a e^{-i\hat t_{clock} h_{\phi}} 
\end{equation}
In \cite{many7} we have suggested to use this form of clock dressing in Cosmology to define the dressed Mukhanov-Sasaki variables \cite{Mukhanov}. As a second remark note that we can use $\hat t_{clock}$ to define a formal $V_s^{clock}$
\begin{equation}\label{clockflow}
V^{clock}_s= e^{is\hat t_{clock}}
\end{equation}
using (\ref{position}) {\it but} with $\hat t$ replaced by $\hat t_{clock}$.

\section{An algebraic model of quantum gravity for closed universes}
From now on we will focus on de Sitter space-times. The main output of the previous discussion is to associate with the type $III_1$ factor $A$ describing the static patch a {\it background independent} type $II_{\infty}$ algebra ${\cal{C}}_{\bar\omega}$. Recall that at this point background independence means independence, up to unitary equivalence, of the algebra ${\cal{C}}_{\bar\omega}$ on the background defined by the state $\phi$ on $A$. As discussed this algebra has a non trivial center ${\cal{Z}}_{\bar\omega}$ generated by the $U_t$ defined in (\ref{momentum}) and consequently is not a factor. In addition we have shown that ${\cal{C}}_{\bar\omega}$ is equipped with a semi finite trace $\tau$. This trace is defined by the reduction of the dominant weight $\bar\omega$ to ${\cal{C}}_{\bar\omega}$. The trace property of the weight $\tau$ follows from the KMS property of the dominant weight $\bar\omega$ \footnote{The trace $\tau$ is, in Taksaki's notation \cite{Takesaki} the weight dual to the dominant weight $\bar\omega$ defined relative to the representation of $A\otimes F_{\infty}= {\cal{C}}_{\bar\omega} \rtimes_{\theta_s} {\mathbb{R}}$.}. The uniqueness of $\bar\omega$ implies the uniqueness of ${\cal{C}}_{\bar\omega}$. This background independent algebra can be represented as the crossed product $A\rtimes_{\sigma_{\phi}}{\mathbb{R}}$ which is by itself independent ( always up to unitary equivalence ) on the background $\phi$.

The Weyl conjugated operator $V_s$ introduced in (\ref{position}) defines the automorphism $\theta_s$ (\ref{auto}) of ${\cal{C}}_{\bar\omega}$ under which the trace $\tau$ scales as
\begin{equation}
\tau(\theta_s(x))=e^{-s}\tau(x)
\end{equation}
for $x\in {\cal{C}}_{\bar\omega}$. This automorphism for elements in ${\cal{Z}}_{\bar\omega}$ defines the {\it flow of weights}.

Next we will use these ingredients to define a toy model of quantum gravity for closed Universes.

\subsection{Finite projections}
Finite projections $P$ in ${\cal{C}}_{\bar\omega}$ are those projections $P^2=P$ having finite trace $\tau(P)$. Recall that equivalent projections $P\sim Q$ are those for which it exists a partial isometry $u$ such that $P=u^*u$ and $Q=uu^*$ and consequently $\tau(P)=\tau(Q)$. Let us denote $[P]$ the class of projections equivalent to $P$. The semi finite trace $\tau$ defined by the dominant weight $\bar\omega$ defines a map from the space of equivalence classes of projections into $[0,\infty]$ that we will denote $\tau([P])$.

Let us now take a finite projection $P$ and let us consider the algebra $P{\cal{C}}_{\bar\omega}P$ acting on $P{\cal{H}}$ with ${\cal{H}}$ the Hilbert space $L^2(\mathbb{R},H_{\phi})$ on which ${\cal{C}}_{\bar\omega}$ is acting. It can be shown that this algebra is type $II_1$. The way to prove it is as follows. Given $P$ let us define a maximal family of {\it equivalent} mutually orthogonal projections $P_i$ such that $P_i \sim P$ and $\sum P_i=1$. Associated to this family we can define a maximal set of infinite unit matrices $e_{i,j}$ satisfying: $e_{i,j}^*=e_{ji}$, $e_{i,j}e_{k,l}=\delta_{j,k}e_{i,l}$ and $\sum_{i}e_{i,i}=1$ and we can identify $P=e_{1,1}$ and the family of equivalent projections by $P_i=e_{i,i}$. This allows us to define a representation of ${\cal{C}}_{\bar\omega}$ as
\begin{equation}\label{repreone}
{\cal{C}}_{\bar\omega} = P{\cal{C}}_{\bar\omega}P\otimes B(l^2({\mathbb{N}}))
\end{equation}
with $B(l^2(\mathbb{N}))$ the algebra generated by the operators $e_{i,j}$ and with $P{\cal{C}}_{\bar\omega}P$ a type $II_1$ algebra. This representation of the type $II_{\infty}$ algebra ${\cal{C}}_{\bar\omega}$ in terms of a type $II_1$ algebra is always possible for any finite projection $P$. 

Note that the former representation only depends on the equivalence class of the finite projection $P$. Thus for each equivalence class $[P]$ we define a type $II_1$ algebra that we can characterize by the concrete value of $\tau([P])$.

\subsection{The maximal entropy state}
Given a particular finite projection $P$ defining the representation (\ref{repreone}) and defining the maximal set of unit matrices $e_{i,j}$ such that $e_{1,1}=P$ we can define the Murray von Neumann coupling constant associated with the type $II_1$ algebra $P{\cal{C}}_{\bar\omega}P$ as acting on $P{\cal{H}}$. This coupling constant $d_{P{\cal{H}}}(P{\cal{C}}_{\bar\omega}P)$ is in the interval $[0,+\infty]$ and consequently there exist a vector state $\xi\in P{\cal{H}}$ that is either cyclic or separating depending if $d_{P{\cal{H}}}(P{\cal{C}}_{\bar\omega}P)$ is smaller or bigger than 1\footnote{See \cite{Jones} prop 10.2.5.}. In any of both cases we can use the vector state $\xi$ to define the type $II_1$ {\it finite} trace on $P{\cal{C}}_{\bar\omega}P$ as
\begin{equation}\label{twoone}
tr(x)=\langle x\xi,\xi\rangle
\end{equation}
for $x\in P{\cal{C}}_{\bar\omega}P$. This {\it vector state $\xi$ that is in $P{\cal{H}}$ is the {\it maximal entropy state} for the type $II_1$ algebra $P{\cal{C}}_{\bar\omega}P$.} 

Using now the identification $P=e_{1,1}$ for the associated family of unit matrices we can define the family of vector states
$\xi_i=e_{i,1}\xi$ and the representation of the semi finite trace on ${\cal{C}}_{\bar\omega}$ as
\begin{equation}\label{repretwo}
\tau(x)= \sum_i\langle x \xi_i,\xi_i\rangle
\end{equation}
Note, using (\ref{repretwo}), that the value $\tau([P])$ associated to the finite projection $P$ satisfies
\begin{equation}\label{reprethree}
\tau(P)= \langle \xi,\xi\rangle = tr(I_P)
\end{equation}
with $tr$ the type $II_1$ trace, $I_P$ the identity in $P{\cal{C}}_{\bar\omega}P$ and  $\xi$ the maximal entropy state defined by this finite projection $P$.

Moreover for the type $II_1$ algebra defined by the finite projection $P$ the trace $tr$ defined in (\ref{twoone}) is {\it normalizable}. Thus for {\it each} finite projection $P$ we can make $tr(I_P)=1$. However, as we will discuss in a moment, we cannot, for two {\it inequivalent} finite projections $P$ and $Q$, to normalize both simultaneously.

Finally for a given finite projection $P$ we can denote the maximal entropy state, whose existence is guaranteed by the finiteness of the corresponding coupling constant, as $\xi_P$. In reality associated to $P$ we define an infinite sequence in $l^2(\mathbb{N},{\cal{H}})$ of vector states in ${\cal{H}}$ defined by $e_{i,1}\xi_P$ for the maximal set of unit matrices $e_{i,j}$ associated to $e_{1,1}=P$. Changing from $P$ to some {\it equivalent} finite projection $Q$ yields to a unitary change of the "basis" vectors $e_{i,1}\xi_P =: \xi_P^i$ into $e_{i,1}\xi_Q=:\xi_Q^i$. 

\subsection{Semiclassical de Sitter space-times}
For any finite projection $P$, and using the same notation as in the previous section, we will define the associated  de Sitter $dS([P])$ space time as defined by:
\begin{itemize}
\item The Hilbert space $P{\cal{H}}$
\item The value of the semi finite trace $\tau(P)$
\item The type $II_1$ algebra $P{\cal{C}}_{\bar\omega}P$
\item The maximal entropy state $\xi_P$
\item A type $II_1$ finite trace on $P{\cal{C}}_{\bar\omega}P$ defined as $tr(x)=\langle x\xi_P,\xi_P\rangle$
\end{itemize}
Any two {\it equivalent} finite projections define equivalent de Sitter space times. Moreover as said above for any concrete de Sitter space time defined in this way we can normalize the trace $tr$. Thus we can map the set of de Sitter space times into the {\it ordered set of equivalence classes} of finite projections in ${\cal{C}}_{\bar\omega}$. Hence we can equally order the different de Sitter space times as $dS([P]) < dS([Q])$ if $\tau([P])<\tau([Q])$.

{\bf Remark 7.1} We could identify the Minkowski limit corresponding to vanishing cosmological constant as the case where we replace the finite projection $P$ by the identity $I$.

{\bf Remark 7.2} It should be stressed that although $\tau(P)$ is {\it finite} and equal for all projections in the equivalence class $[P]$ the Hilbert space $P{\cal{H}}$ on which the type $II_1$ factor $P{\cal{C}}_{\bar\omega}P$ is acting is {\it not} finite dimensional. You can prove the existence in $P{\cal{C}}_{\bar\omega}P$ of subfactors isomorphic to $M_n({\mathbb{C}})$ for any $n\in {\mathbb{N}}$. Intuitively we can think of $M_n(\mathbb{C})$ as the algebra representing $n$ q-bits \footnote{See \cite{Jones} 6.1.18.}. 

\subsubsection{The affiliated density matrices}
Let us now consider a given finite projection $P$ and let us examine what the  different states $\Psi$ in the Hilbert space $P{\cal{H}}$ describe. 

The vector state $\Psi$ has an {\it affiliated self adjoint density matrix} $h_{\Psi}$ \footnote{Affiliated to the type $II_1$ algebra $P{\cal{C}}_{\bar\omega}P$.} defined by\footnote{See \cite{Connes1} I.0 \cite{Longo} and discussion in section 4 of this note.}
\begin{equation}
\langle h_{\Psi} \cdot \xi_P,\xi_P\rangle = \langle \cdot \Psi,\Psi\rangle
\end{equation}
for $\cdot$ representing any element in $P{\cal{C}}_{\bar\omega}P$. Moreover we can define a von Neumann entropy \footnote{See \cite{Longo}.} associated to $\rho_{\Psi} =:h_{\Psi}$ as
\begin{equation}\label{vN}
S_{VN}(\rho_{\Psi})= -\langle \xi_P, \rho_{\Psi}\ln\rho_{\Psi} \xi_P\rangle
\end{equation}
Obviously $\rho_{\xi_P}=1$ if we normalize $\langle\xi_P,\xi_P\rangle=1$. Physically if we think of $\Psi$ as representing a state on the de Sitter space associated to $P$, the vN entropy (\ref{vN}) accounts for the reduction of the Gibbons Hawking entropy of pure de Sitter due to whatever is the perturbation inside, represented by the state $\Psi$. Recall that $\Psi$ is in $P{\cal{H}}$. The probability amplitude to create the fluctuation represented by $\Psi$ is suppressed by $e^{S_{VN}(\rho_{\Psi})}$ \footnote{The underlying physics is well known \cite{Banks1,Banks2,Banks3,Banks4,Banks5,Banks6,Banks7,Banks8,Banks9,Banks10,Banks11} and it is described in \cite{Witten1}.}. Note that since the affiliated density matrix $h_{\xi_P}=1$ the VN entropy defined in (\ref{vN}) agrees with the relative entropy.

{\bf Remark 7.3} Note that the set of projections in the type $II_1$ factor $P{\cal{C}}_{\bar\omega}P$ is ordered. That implies, by Zorn lemma, the existence of a  maximal element that is $I_P$. This is the reason to call the state $\xi_P$ the maximal entropy state. In addition we can find projections $p\in P{\cal{C}}_{\bar\omega}P$ with arbitrarily small value of $tr(p)$\footnote{The physics meaning of this fact deserves further attention.}.

\subsection{Entropy of $dS([P])$}
Now we need to understand what is the meaning of $\tau([P])$ for the space time $dS([P])$. In order to do that let us consider two inequivalent finite projections $P$ and $Q$ that should correspond, in the present scheme, to two different dS space times. For both projections we can define two sequences $\xi_P^{i}$ and $\xi_Q^i$ in $l^2(\mathbb{N},{\cal{H}})$. These two sequences define two {\it normal states} let us say $\varphi_P$ and $\varphi_Q$ of ${\cal{C}}_{\bar\omega}$. In principle associated to these two normal states we can define the affiliated operators $h_{\varphi_P}$ and $h_{\varphi_Q}$ and to define the corresponding VN entropies as well as the relative entropy between both dS space times. Generically in this case where we are using the semi finite trace $\tau$ in order to define the affiliated operators, we cannot define a von Neumann entropy as $-\tau(h_{\varphi_P}\ln h_{\varphi_P})$. The reason is that $\tau$ is only defined as a trace on the positive operators in ${\cal{C}}_{\bar\omega}^{+}$ and we need to work with those affiliated operators such that $\ln h_{\varphi_P}$ is in ${\cal{C}}_{\bar\omega}^{+}$ ( see \cite{Longo} ). 

For one finite projection $P$ that we can take as reference we can always normalize $\tau(P)=1$ {\it but} we cannot normalize all the inequivalent projections but just the one that we take as reference. In these conditions we can assign to any projection $Q$ inequivalent to $P$ a formal entropy defined by
\begin{equation}\label{basic1}
S(dS[Q]) = -\ln \tau([Q])
\end{equation}
with $\tau([Q]) \in [0,\infty[$. This is the quantity we will put in correspondence with the generalized Bekenstein Hawking Gibbons entropy of the de Sitter space time associated to $[Q]$. The reason this is a {\it generalized} entropy is because the semifinite trace $\tau$ was defined by the dominant weight $\bar\omega$ on $A\otimes F_{\infty}$ with $F_{\infty}$ formally associated to an external observer.

{\bf Remark 7.4} As stressed in Remark 7.2 above the finite value of $\tau(P)$ for a given de Sitter class $[P]$ cannot be ( holographically) interpreted as defining a representation of $P{\cal{H}}$ as the Hilbert space of $\tau(P)$ q-bits. This would be the case for type $I$ algebras but not for the type $II_1$ algebra $P{\cal{C}}_{\bar\omega}P$ \footnote{Note that $\tau(P)$ is finite after we normalize with respect to a reference projection $[P]$ with $\tau([P])=1$. This should be taken into account in any microscopic attempt to interpret the finiteness of (\ref{basic1}).}.
\subsubsection{Quantum gravity amplitudes}
The way to define {\it quantum gravity} amplitudes as quantum transitions between different dS space times becomes possible when the two inequivalent finite projections $P$ and $Q$ used to define two different de Sitter space times are {\it not mutually orthogonal}. Recall that for each finite projection $P$ we have defined a maximal set of equivalent projections $P_i$ mutually orthogonal with $P_i\sim P$. The maximal entropy state $\xi_P\in P{\cal{H}}$ is associated with the sequence of states $\xi_P^i$ in $l^2(\mathbb{N},{\cal{H}})$. Let us now consider an inequivalent finite projection $Q$. This means we are considering a different de Sitter space $dS([Q])$ characterized by $\tau([Q])$ and with a sequence associated to the maximal entropy state $\xi_Q$. The simplest way to define a quantum gravity amplitude between these two space times is either as the amplitude
\begin{equation}
\langle \xi_P,\xi_Q\rangle
\end{equation}
defined in ${\cal{H}}$ or more precisely the scalar product in $l^2({\mathbb{N}},{\cal{H}})$ for the two {\it maximal entropy} sequences defined by $\xi_P^i$ and $\xi_Q^i$. 

Heuristically we can think these amplitudes as follows. For a given $P$ i.e for a given de Sitter we define the "basis" of vectors $\xi_P^i$. Now given a different de Sitter defined by an inequivalent finite projection $Q$ we identify the maximal entropy state $\xi_Q$ and we define the {\it quantum gravity} amplitudes $c_i(P,Q)$ as $\langle \xi_Q,\xi_P^i\rangle$. Thus we get
\begin{equation}\label{basic2}
c_i(P,Q)= \langle \xi_Q,e_{i,1}^{P}\xi_P\rangle
\end{equation}
with $e_{i,1}^P$ defined the maximal set of unit matrices associated to $P$ i.e. with $e_{1,1}=P$ \footnote{The physics meaning of the label $i$ deserves some further clarification.}.
\subsection{The dS flow}
Let us now consider the one parameter automorphism $\theta_s$ of ${\cal{C}}_{\bar\omega}$. The crossed product of ${\cal{C}}_{\bar\omega}$ by this automorphism is by Takesaki duality isomorphic to the algebra $A\otimes F_{\infty}$. Let us now consider a finite projection $P$ defining the particular de Sitter space $dS([P])$. Under the action of $\theta_s$ we get a new finite projection defined as
\begin{equation}
P_s= \theta_s P\theta_s^{*}
\end{equation}
with 
\begin{equation}\label{flow}
\tau(P_s)= e^{-s} \tau(P)
\end{equation}
Thus if we identify $\tau(P)$ as representing the generalized Gibbons Hawking entropy of the $dS([P])$ space time, then the automorphism $\theta_s$ defines a flow of $dS([P])$ into a $dS([P_s])$. Recall that (\ref{flow}) comes from the definition of the dominant weight $\bar\omega$ as $\phi\otimes \omega$ with $\omega$ the weight on $F_{\infty}$ defined by $Tr(\cdot,\rho)$ with $\rho$ the generator of translations in $L^2(\mathbb{R})$ and, as usual, $\cdot$ representing any element in $F_{\infty}$. Taking now into account that $\theta_s$ is defined by $V_s$ (\ref{position}) and using the Weyl CCR we get (\ref{flow}). It is important here, from the point of view of physics, to keep in mind that the Weyl CCR are used here in the convention  $\hbar=1$. In reality hidden in (\ref{flow}) there exist a $\hbar$ that we have conventionally defined as one. 

Once we reintroduce explicitely $\hbar$ we observe the {\it quantum meaning} of the de Sitter flow defined by (\ref{flow}). 

Now we can come back to the question of {\it normalization}. For a given finite projection $P$ we can use the automorphism $\theta_s$ to normalize $\tau(P)=1$. This is equivalent to the normalization of the type $II_1$ trace $tr$ defined using the state $\xi_P$. When we have two inequivalent finite projections $P$ and $Q$ we cannot use the one parameter automorphism $\theta_s$ to normalize $\tau(P)$ {\it and} $\tau(Q)$. Thus the {\it relative} values of $\tau(P)$ for inequivalent finite projections have physical meaning.

\subsection{The HH weight}
Intuitively we could try to define a Hartle Hawking \footnote{For a review on HH and a good collection of references see \cite{Anninos}.} weight $\bar\omega^{HH}$ as defining the so called no boundary quantum gravity partition function. For the case of de Sitter space times we could define
\begin{equation}\label{partition}
Z=\sum_{\alpha} e^{S(\alpha)}
\end{equation}
with the sum over different de Sitter space times and with $S(\alpha)$ representing the corresponding Gibbons Hawking generalized entropy. Let us look for the weight $\bar\omega^{HH}$ by imposing
\begin{equation}
\bar\omega^{HH}(I)=Z
\end{equation}
In our former approach we have identified each dS space time with the equivalence class of  a finite projection, so the sum in (\ref{partition}) should be interpreted as a sum over the {\it whole ordered set of equivalence classes of finite projections} that we will represent as $[P]_{\alpha}$. Moreover we have interpreted $e^{S(\alpha)}$ in $Z$ as $\tau([P]_{\alpha})$. Finally we have introduced the dS flow induced by the automorphism $\theta_s$ to flow from a finite projection with trace $\tau(P)$ into a new projection $P_s$ with trace $e^{-s}\tau(P)$. Thus we could try to represent the integral in $Z$ over different de Sitter space times as an integral over the action of $\theta_s$ defining
\begin{equation}\label{basic3}
Z=: \bar\omega^{HH}(I)= \int \tau([P_s]) ds
\end{equation}
or equivalently $\bar\omega^{HH}(\cdot)=\int \bar\omega(P_s \cdot P_s) ds$.

In summary the quantum gravity approach to closed Universes presented above is defined by (\ref{basic1}),(\ref{basic2}) and (\ref{basic3}).

{\bf Remark 7.5} If we are doing Cosmology in a particular semiclassical de Sitter  defined by the equivalence class of a finite projection $P$ we will be interested in defining physical fluctuations using operators of type $e_{1,1}d(a)e_{1,1}$ for $d$ the dressing operator, $a$ an element in $A$ and $e_{1,1}=P$. Following our former discussion on clocks we could think that there exist a clock dressing $d_{clock}$ such that $d_{clock}(a)=
e_{1,1}d(a)e_{1,1}$. Cosmological observables, as for instance curvature power spectrum
should be defined by $\langle\xi_P,d_{clock}(a)(d_{clock}(a))^*\xi_P\rangle$. As discussed in \cite{many7} if we use as clock coordinate the inflaton vev the former dressing will implements the familiar dependence of the power spectrum on the slow roll parameter.

{\bf Remark 7.6} Again in the context of Cosmology we could try to define a {\it Cosmological dS flow} using instead of the automorphism $\theta_s$ defined by $V_s$ the {\it clock} automorphism (\ref{clockflow}) defined by $\hat t_{clock}$ as $V^{clock}_s$ using a sort of cosmological clock as the one defined in \cite{many7}. 

\subsubsection{Relation to the HH no boundary partition function}
In order to make contact with the suggestion in \cite{Witten0} let us recall the role of the observer Hamiltonian $m+q$ ( in the notation of \cite{Witten0} and \cite{Witten1}) in the definition of the weight $\bar\omega$. In \cite{Witten1} the {\it observer} is equipped with a hamiltonian $m+q$ that can be identified with a physical clock once we project on the positive part of the spectrum of $m+q$. So formally the observer is defined by a finite projection $P$ in the centralizer ${\cal{C}}_{\bar\omega}$. The way to define physically this projection is, as said, introducing the hamiltonian $m+q$ with $m+q$ positive. Note that $q$ is defining a hamiltonian {\it operator} acting on the {\it observer} Hilbert space $L^2(\mathbb{R})$. In the notation used in this note $m+q$ is the hamiltonian $H_r$ defined in (\ref{observerweight}). Different observer {\it positive} hamiltonians correspond to different {\it positive} $q$'s and consequently to different finite projections. In particular for the finite projection $P$ corresponding to impose $q \geq 0$ the maximal entropy state $\xi_P$ is the one used in \cite{Witten1}. Let us denote $P_q$ this finite projection.

Now the automorphism $\theta_s$ acts on the observer hamiltonian according to the CCR described in section 3.3 i.e. as {\it translations} of $q$. Hence the {\it no boundary} partition function defined as the integral over $q$ of $\langle \xi_{P_q},\xi_{P_q}\rangle$ is replaced by (\ref{basic3}) as 
\begin{equation}
Z= \int \bar\omega([P_s]) ds
\end{equation}
where we integrate over the full set of {\it equivalence classes of finite projections} each one represented by the value of the finite trace $\tau([P_s])$ defined as the weight dual to $\bar\omega$.

\section{Acknowledgments}
This work was supported by grants SEV-2016-0597, FPA2015-65480-P and PGC2018-095976-B-C21.

\end{document}